\documentclass[sigconf]{acmart}

\copyrightyear{2026}
\acmYear{2026}
\setcopyright{cc}
\setcctype{by}
\acmConference[KDD 2026] {Proceedings of the 32nd ACM SIGKDD Conference on Knowledge Discovery and Data Mining V.2}{August 9--13, 2026}{Jeju Island, Republic of Korea.}
\acmBooktitle{Proceedings of the 32nd ACM SIGKDD Conference on Knowledge Discovery and Data Mining V.2 (KDD 2026), August 9--13, 2026, Jeju Island, Republic of Korea}
\acmISBN{979-8-4007-2259-2/2026/08}
\acmDOI{10.1145/3770855.3818074}

\AtBeginDocument{%
  }

\usepackage{colortbl}
\usepackage{booktabs}   
\usepackage{multirow}   
\usepackage{graphicx}   
\usepackage{algorithm}
\usepackage{algpseudocode}
\usepackage{float}
\usepackage[table]{xcolor} 
\usepackage{xcolor}
\usepackage[most]{tcolorbox}
\usepackage{subcaption} 
\usepackage{tabularx}

\begin{document}

\title{MemGraphRAG: Memory-based Multi-Agent System for Graph Retrieval-Augmented Generation}







\author{Chuanjie Wu}
\authornote{Contributed equally.}
\email{wuchuanjie@stu.xmu.edu.cn}
\affiliation{%
  \institution{Xiamen University\textsuperscript{1, 2}}
  \city{Xiamen}
  \country{China}
}
\author{Zhishang Xiang}
\email{xiangzhishang@stu.xmu.edu.cn}
\authornotemark[1]
\affiliation{%
  \institution{Xiamen University\textsuperscript{2, 3}}
  \city{Xiamen}
  \country{China}
}
\author{Yunbo Tang}
\email{tangyunbo@stu.xmu.edu.cn}
\affiliation{%
  \institution{Xiamen University\textsuperscript{1}}
  \city{Xiamen}
  \country{China}
}
\author{Zerui Chen}
\email{chenzerui1@stu.xmu.edu.cn}
\affiliation{%
  \institution{Xiamen University\textsuperscript{1}}
  \city{Xiamen}
  \country{China}
}
\author{Qinggang Zhang}
\email{qinggangzhang@jlu.edu.cn}
\authornotemark[2]
\affiliation{%
  \institution{Jilin University}
  \city{Changchun}
  \country{China}
}
\author{Jinsong Su}
\authornote{Corresponding author.}
\email{jssu@xmu.edu.cn}
\affiliation{%
  \institution{Xiamen University\textsuperscript{1, 2, 3}}
  \city{Xiamen}
  \country{China}
}
\begin{abstract}

\footnotetext[1]{School of Informatics}
\footnotetext[2]{Key Laboratory of Digital Protection and Intelligent Processing of Intangible Cultural Heritage of Fujian and Taiwan,Ministry of Culture and Tourism}
\footnotetext[3]{Institute of Artificial Intelligence}


  Retrieval-Augmented Generation (RAG) has become an essential method for mitigating hallucinations in Large Language Models (LLMs) by leveraging external knowledge. Although effective for simple queries, traditional RAG struggles with large-scale, unstructured corpora where information is highly fragmented. Graph-based RAG (GraphRAG) incorporates knowledge graphs to capture structural relationships, enabling more comprehensive retrieval for complex reasoning. However, existing GraphRAG methods rely on isolated, fragment-level extraction for graph construction, lacking a global perspective on the whole corpus. As a result, these methods frequently lead to thematically inconsistent, logically conflicting, and structurally fragmented graphs that degrade retrieval performance. In this paper, we propose MemGraphRAG, a novel framework that introduces a memory-based multi-agent system to ensure high-quality graph construction. Specifically, MemGraphRAG employs a collaborative society of agents supported by shared memory, which provides a unified global context throughout the extraction process. This mechanism allows agents to dynamically resolve logical conflicts and maintain structural connectivity throughout the corpus. Furthermore, we propose a memory-aware hierarchical retrieval algorithm tailored for the constructed graph. Extensive experiments on multiple benchmarks demonstrate that MemGraphRAG outperforms the state-of-the-art baseline models with comparable efficiency. Our code is available at \textcolor{blue}{\url{https://github.com/XMUDeepLIT/MemGraphRAG}}.
\end{abstract}




\maketitle
\begin{figure}[t] 
\label{fig:introduction}
    \centering
    
    \includegraphics[width=0.48\textwidth]{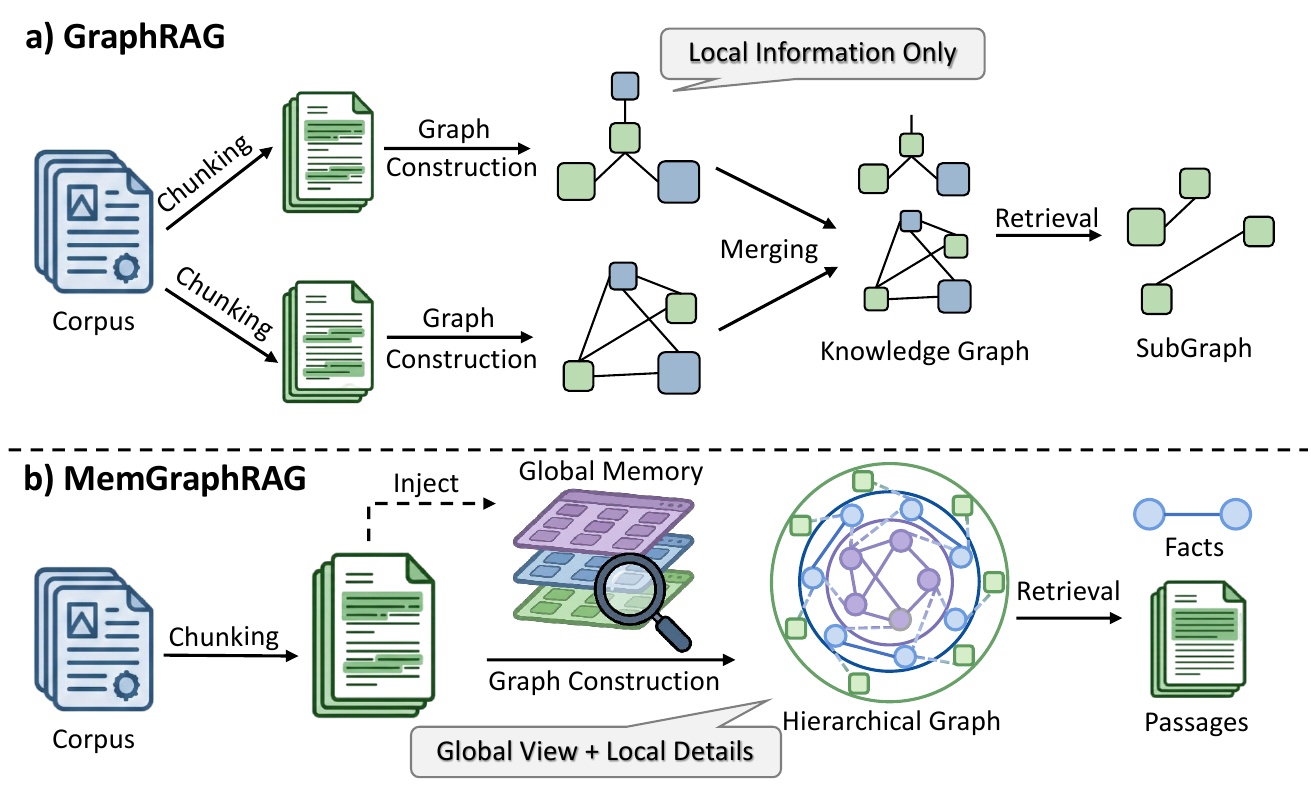} 
    \caption{Comparison between existing GraphRAG and MemGraphRAG. Exiting GraphRAG performs isolated chunk-level extraction without the global view, resulting in a noisy and inconsistent indexing graph. While MemGraphRAG incorporates global memory to ensure global consistency.}
    
    \label{fig:introduction}
    \vspace{-2mm}
\end{figure}
\section{Introduction}

Recently, Retrieval-Augmented Generation (RAG) effectively extends the capabilities of Large Language Models (LLMs) by leveraging external knowledge~\cite{gao2023retrieval,lewis2020retrieval,zhang2025faithfulrag}. However, existing RAG systems suffer from critical challenges in real-world scenarios. This is due to the unstructured and heterogeneous nature of large-scale corpora, where relevant information is often sparsely distributed. The contexts retrieved by RAG systems are often noisy and lack structural coherence. Although recent methods attempt to segment documents into smaller chunks for efficient indexing~\cite{borgeaud2022improving,izacard2023atlas,jiang2023active}, this strategy disrupts long-range dependencies and loses critical contextual details. As a result, the retrieved contexts are often incoherent or insufficient for complex reasoning tasks~\cite{han2024retrieval,zhang2025survey}.

To address these limitations, Graph Retrieval-Augmented Generation (GraphRAG)~\cite{zhang2025survey,procko2024graph,xiang2025use} has emerged as a powerful paradigm, leveraging external structured graphs to model the hierarchical structure of background knowledge~\cite{han2024retrieval,yang2026graph}.
Early efforts, such as RAPTOR~\cite{sarthi2024raptor} and Microsoft’s GraphRAG~\cite{edge2024local}, organize knowledge through recursive summarization and community-level abstractions to support coarse-to-fine retrieval, thereby facilitating comprehensive response generation. 
Subsequent methods, including GFM-RAG~\cite{luo2025gfm}, G-Retriever~\cite{he2024g}, and LightRAG~\cite{guo2024lightrag}, further incorporate specialized retrieval mechanisms and learning objectives to improve multi-hop generalization, scalability, and efficiency. 
Most recently, HippoRAG~\cite{hipporag}and its enhancement HippoRAG2~\cite{gutiérrez2025hipporag2} have drawn inspiration from cognitive associative memory, utilizing algorithms such as Personalized PageRank to simulate multi-hop reasoning pathways. 
These strategies demonstrate the potential of graph-baed retrieval in addressing the core limitations of traditional RAG.


\begin{figure}[t]
\centering  
\begin{subfigure}[t]{0.47\linewidth}    
\centering    
\includegraphics[width=\linewidth]{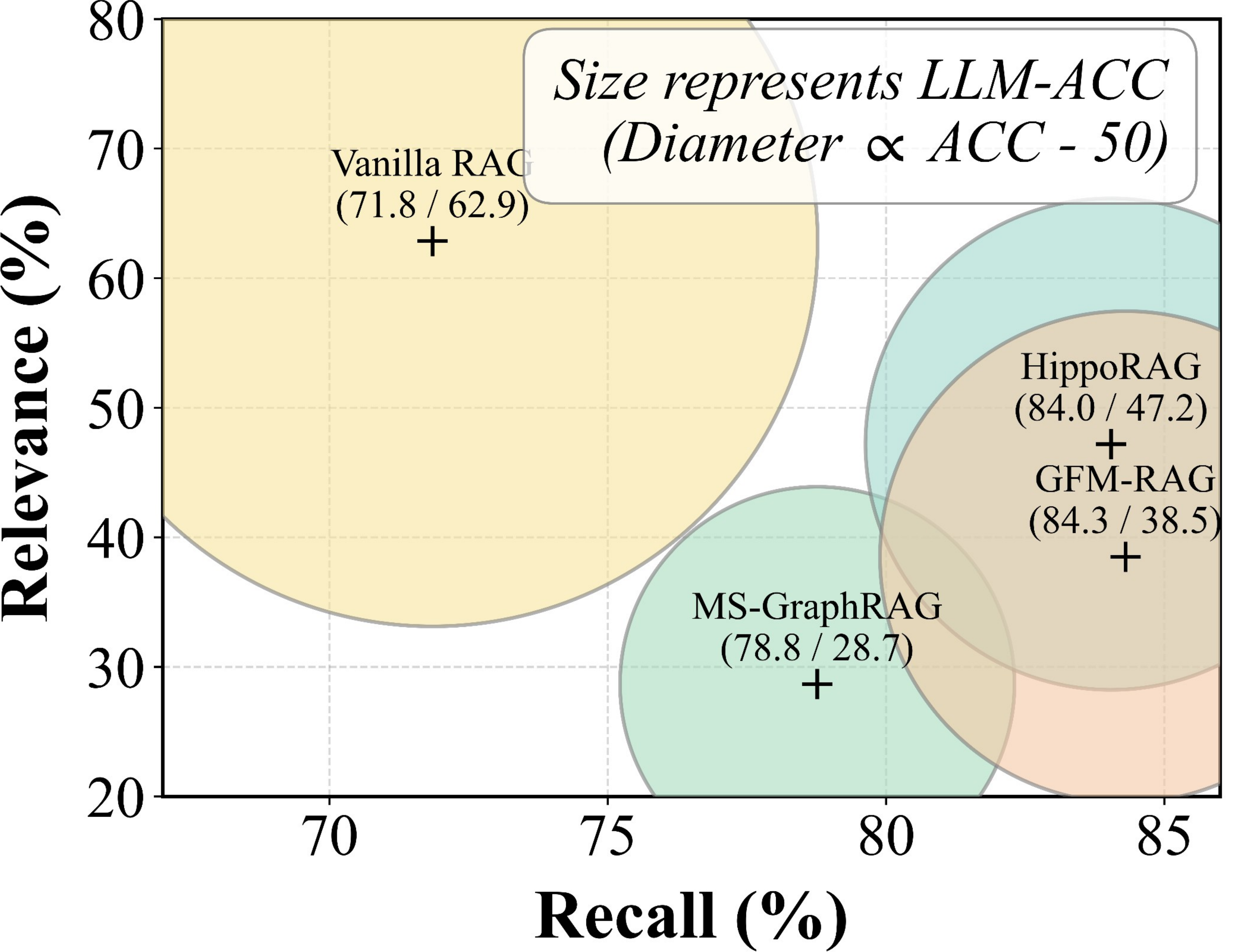}    
\end{subfigure}  
\begin{subfigure}[t]{0.49\linewidth}    
\centering    
\includegraphics[width=\linewidth]{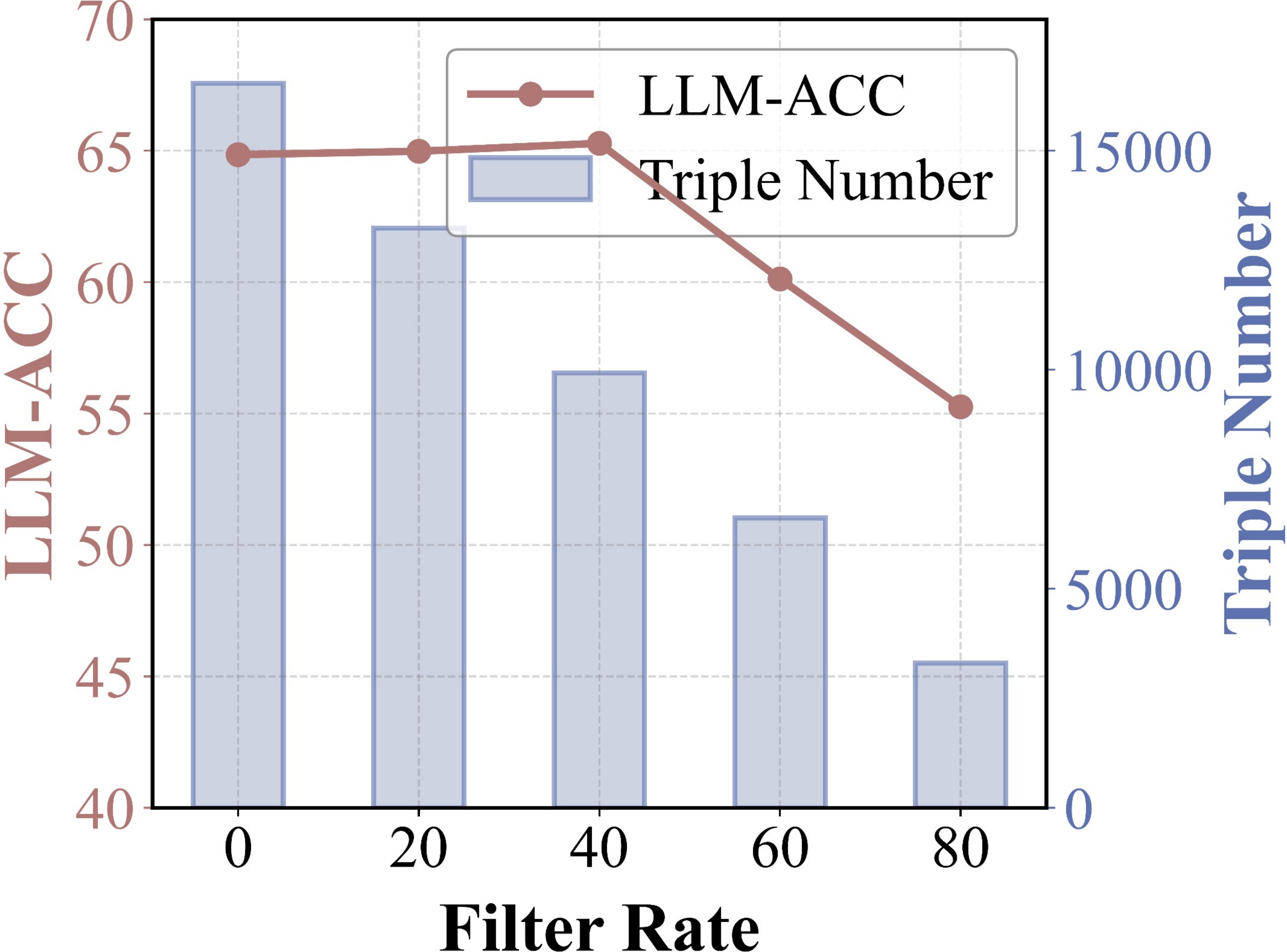}    
\end{subfigure}
\vspace{-4mm}
\caption{\textbf{(Left)} Evaluation of representative RAG and GraphRAG systems. The radius reflecting performance of each systems. \textit{Relevance} measuring context relevance to the query. \textit{Recall} measuring whether sufficient evidence is covered. \textbf{(Right)} Impact of removing irrelevant triples based on schema frequency on the final performance (LLM-ACC).}
\label{fig:pre1}
\vspace{-5mm}
\end{figure}

However, contrary to their theoretical advantages, GraphRAG systems frequently underperform naive RAG systems in many real-world applications~\cite{han2025rag,zhou2025depth,xiang2025use,zhuang2025linearrag}. This performance decline is primarily due to the low quality of automatically constructed knowledge graphs~\cite{xiang2025use,zhuang2025linearrag}. Although graph-based retrieval enhances relevant knowledge recall, errors in graph construction introduce substantial noise into the retrieved contexts simultaneously. Fundamentally, these challenges persist because existing pipelines typically derive knowledge from isolated local segments, lacking a global perspective on the previously processed context. This isolation leads to three critical deficiencies that undermine graph quality: (i) \textbf{thematic irrelevance}: extracted triples are often irrelevant to the central theme, introducing meaningless facts.
(ii) \textbf{logical inconsistency}: contradictory facts may emerge within a single subgraph, compromising semantic coherence. 
(iii) \textbf{structural fragmentation}: the built graphs often suffer from fragmentation issues, where the isolated nodes and disconnected components weaken the core advantage of the knowledge graph in supporting global comprehension and multi-hop reasoning. 

While some recent studies attempt to improve graph quality before extraction by filtering triples using predefined schema~\cite{sharma2024og,dong2025youtugraphrag}, these approaches suffer from limited generalization and high manual costs. 
Other efforts seek to improve graph quality through bottom-up clustering-based community summarization~\cite{edge2024local, gutiérrez2025hipporag2, wang2025archrag} or topic modeling~\cite{sarthi2024raptor}. Nevertheless these unsupervised approaches remain susceptible to error propagation, because inaccuracies in entity relations tend to be amplified at high-level  summaries.



 To address this, we revisit the pipeline of existing GraphRAG systems and propose a \underline{\textbf{Mem}}ory-Based Multi-Agent Framework for \underline{\textbf{Graph}} \underline{\textbf{R}}etrieval-\underline{\textbf{A}}ugmented \underline{\textbf{G}}eneration (\textbf{MemGraphRAG}). Specifically, MemGraphRAG employs a collaborative society of agents supported by a novel Three-Layer
Global Memory. This shared memory structure serves as a unified
knowledge repository, providing a global perspective that enables agents to dynamically coordinate
the extraction process, resolve conflicts upon detection, and integrate fragmented information across
the entire corpus.
To summarize, our contributions are listed as follows:
\begin{itemize}
    \item We identify the root cause of performance degradation in existing GraphRAG systems: the reliance on isolated local extraction. We demonstrate how this lack of global context inevitably leads to three critical deficiencies: thematic irrelevance, logical inconsistency, and structural fragmentation.
    
    \item 
    We propose MemGraphRAG, which introduces a memory-based multi-agent system into graph construction. The shared memory not only maintains global thematic consistency to prevent irrelevance and fragmentation, but also provides grounded evidence to resolve local logical inconsistencies. Besides, we propose a memory-aware hierarchical retrieval algorithm tailored for the constructed graph.
    \item We conduct extensive experiments on four benchmark datasets, demonstrating that MemGraphRAG consistently outperforms state-of-the-art baselines in terms of graph quality, retrieval quality and generation accuracy, validating its practicality for real-world applications.
    
\end{itemize}

\section{Problem Statement}
To facilitate subsequent discussion, we first introduce key definitions for the knowledge representation, and then present the complete problem formulation of GraphRAG.
\subsection{Key Definitions}
We first provide formal definitions for the core components of our knowledge representation:

(i) \textbf{type ($t$) and entity ($e$)}: a type $t$  (e.g., \textit{person}) denotes an abstract category, while an entity $e$ (e.g., \textit{Einstein}) is a concrete instance. Formally, a typing function $\phi$ assigns each entity to its specific type, denoted as $\phi(e)=t$. 

(ii) \textbf{schema ($s$) and fact ($f$)}: a schema $s=(t_h, r, t_t)$ (e.g., (\textit{person}, \textit{born\_in}, \textit{country})) specifies a logical constraint. $t_h, t_t$ represent the head and tail types, respectively, $r$ denotes a semantic relation. Based on this structure, a fact $f=(e_h, r, e_t)$ (e.g., (\textit{Einstein}, \textit{born\_in}, \textit{Germany})) is a concrete instantiation of a schema, where $e_h, e_t$ represent the head and tail entity.

(iii) \textbf{ontology ($\mathcal{O}$)}: the ontology $\mathcal{O}$ is defined as the collection of all valid schemas, denoted as $\mathcal{O}=\{s_1,\dots,s_{|\mathcal{O}|}\}$. It includes the theme and logical rules of the whole knowledge graph.

(iv) \textbf{passage ($p$)}: a passage $p$ denotes the specific text segment acting as the source of the extracted information. We define a function $\psi(f) = p$ to trace each fact $f$ back to its origin.

Detailed definitions are provided in Appendix~\ref{appendix:notation}.

\subsection{Problem Formulation} We formally formulate the task of GraphRAG as a unified framework composed of two distinct phases: 

(i) \textbf{Offline Graph Structure Construction}. Given a corpus of unstructured documents $\mathcal{D}=\{d_1, d_2, ..., d_{|\mathcal{D}|}\}$, the primary objective is to transform raw text into a structured graph $\mathcal{G}=(\mathcal{V}, \mathcal{E})$. In our framework, the vertex set $\mathcal{V}$ is heterogeneous, comprising entities, types, and passages ($\mathcal{V} = \mathcal{V}_e \cup \mathcal{V}_t \cup \mathcal{V}_p$) ,  and the edge set $\mathcal{E}$ encodes the semantic dependencies between them. Formally, this construction process is formalized as
\begin{equation}\small
\mathcal{G} = \texttt{GraphConstructor}(\mathcal{D}) 
\end{equation}
where \texttt{GraphConstructor(*)} maps the unstructured corpus to a semantic graph topology, facilitating the efficient navigation from abstract concepts to concrete evidence.

(ii) \textbf{Online Graph-Enhanced Retrieval and Reasoning}. Based on the constructed graph $\mathcal{G}$, the system processes a user query $q$ to generate a final answer $a$. Unlike extracting isolated text segments, this phase involves identifying optimal reasoning paths within the graph to curate a structured context. The process is formulated as
\begin{equation}\small
a = \texttt{LLM}(\texttt{Retriever}(q, \mathcal{G})) 
\end{equation}
where \texttt{Retriever(*)} identifies the most relevant graph elements (subgraphs) to support grounded answer generation.



    
    

\begin{figure}[t] 
    \centering  
    \includegraphics[width=\linewidth]{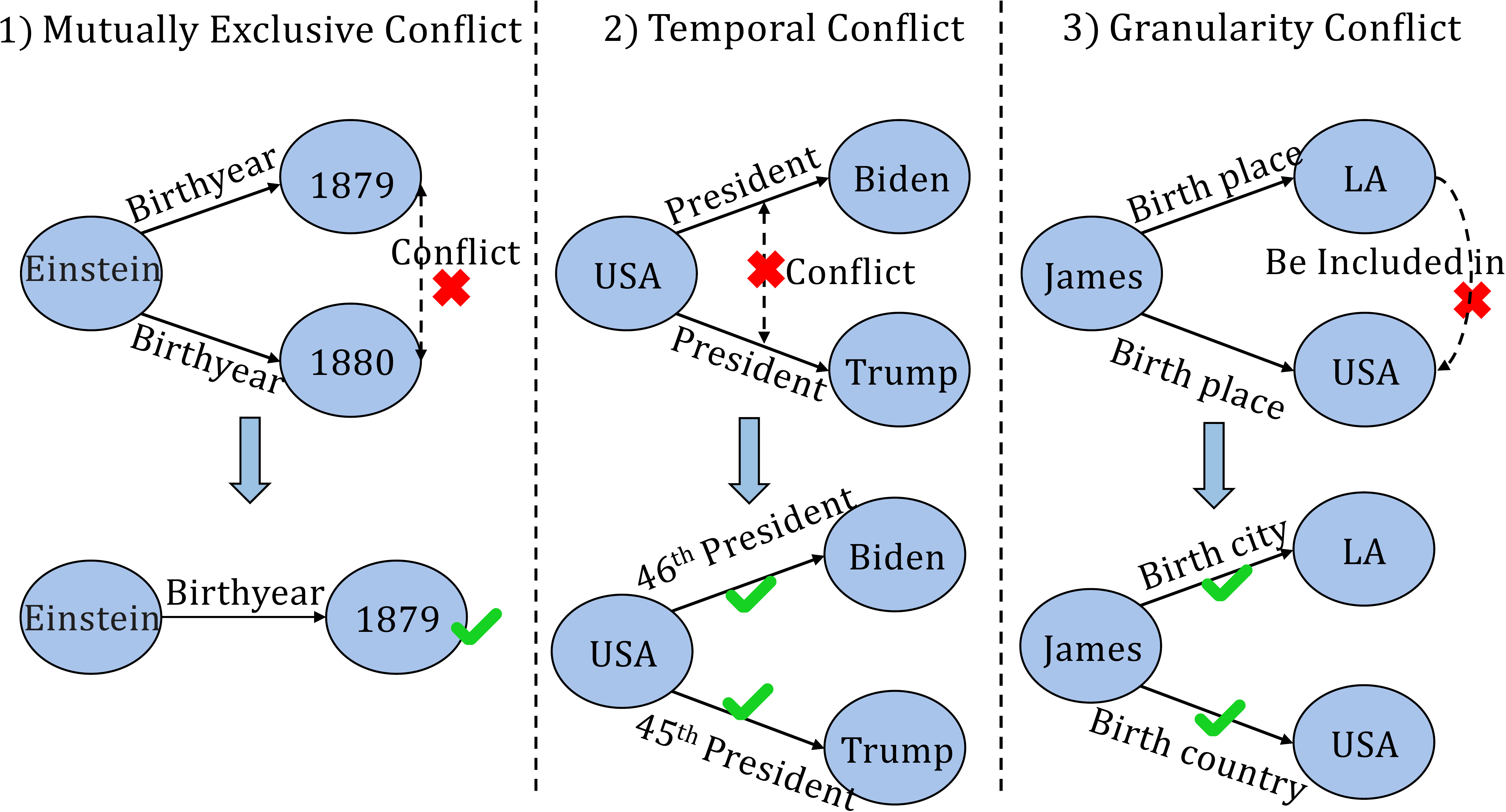} 
    \caption{Illustration of three conflict types in extracted graphs: 1) Mutually Exclusive Conflict from logically incompatible facts, 2) Temporal Conflict caused by missing temporal grounding for time-varying states, and 3) Granularity Conflict arising from inconsistent abstraction levels for the same entity or concept. Details are in Table~\ref{tab:knowledge_conflicts} in Appendix~\ref{app:pre}}
    \label{fig:conflict}
    \vspace{-5mm}
\end{figure}

\section{Preliminary Study}\label{sec:preliminary}
Although knowledge graphs can model complex dependencies, recent benchmarks show that advanced GraphRAG systems may underperform naive RAG in real-world QA tasks~\cite{xiang2025use,zhuang2025linearrag}. To investigate this issue, we conduct two preliminary studies to analyze how automatic graph construction affects retrieval quality and downstream generation.

\subsection{Performance Degradation}

We first compare Vanilla RAG with recent GraphRAG systems (MS-GraphRAG, HippoRAG, and GFM-RAG) in the G-Medical dataset\cite{xiang2025use}. As shown in Figure~\ref{fig:pre1}(a), GraphRAG methods achieve higher retrieval Recall (e.g., GFM-RAG: 84.3\% vs. RAG: 71.8\%), but suffer a substantial drop in Relevance (38.5\% vs. 62.9\%), leading to noisier contexts and lower generation accuracy. These results indicate that existing GraphRAG pipelines often expand the retrieval coverage at the cost of introducing excessive irrelevant information, which ultimately harms the QA performance.

\subsection{Error Analysis}
To further investigate why graph construction introduces noise and conflicts, we hypothesize that the root cause lies in the isolated local extraction paradigm adopted by most baselines. Without a persistent global memory, extraction LLMs process document chunks independently, which leads to systematic issues in graph quality. Specifically, we summarize the major failure modes as follows: 

\textbf{Thematic Irrelevance.} Without a global view of the corpus theme, local extraction tends to introduce off-topic triples. To quantify this effect, we conduct a filtering experiment (Figure~\ref{fig:pre1}(b)) that removes triples based on schema frequency. Interestingly, filtering out 40\% of low-frequency triples slightly improves accuracy (65.28\% vs. 64.85\%), suggesting that a large fraction of extracted triples are thematically irrelevant noise.

\textbf{Logical Inconsistency.} Independent extraction also introduces semantic contradictions into the merged graph. As illustrated in Figure~\ref{fig:conflict}, we observe mutually exclusive conflicts, temporal conflicts, and granularity conflicts, which create inconsistent reasoning paths and confuse downstream retrieval. More conflict analyses are provided in Appendix~\ref{app:pre}.

\textbf{Structural Fragmentation.} Due to missing global coreference resolution and schema alignment, key entities are often duplicated or scattered across disconnected subgraphs. This fragmentation prevents effective multi-hop traversal and reduces the usefulness of the graph for global reasoning.

\subsection{Discussion} Current GraphRAG systems exhibit two fundamental limitations. \textbf{First, existing GraphRAG systems exhibit a fundamental trade-off between recall and relevance.} Although graph expansion improves coverage, it often retrieves irrelevant evidence that overwhelms the LLM and degrades generation accuracy. \textbf{Second, current GraphRAG systems lack a global memory mechanism during graph construction.} Most systems rely on isolated local extraction, processing document chunks independently without maintaining a persistent global state. As a result, the constructed graph fails to preserve thematic coherence and resolve cross-document conflicts, leading to \textit{thematic irrelevance}, \textit{logical inconsistency}, and \textit{structural fragmentation} in downstream retrieval and reasoning processes.

\section{Our Framework}
\begin{figure*}[t] 
    \centering
    \includegraphics[width=1.\textwidth]{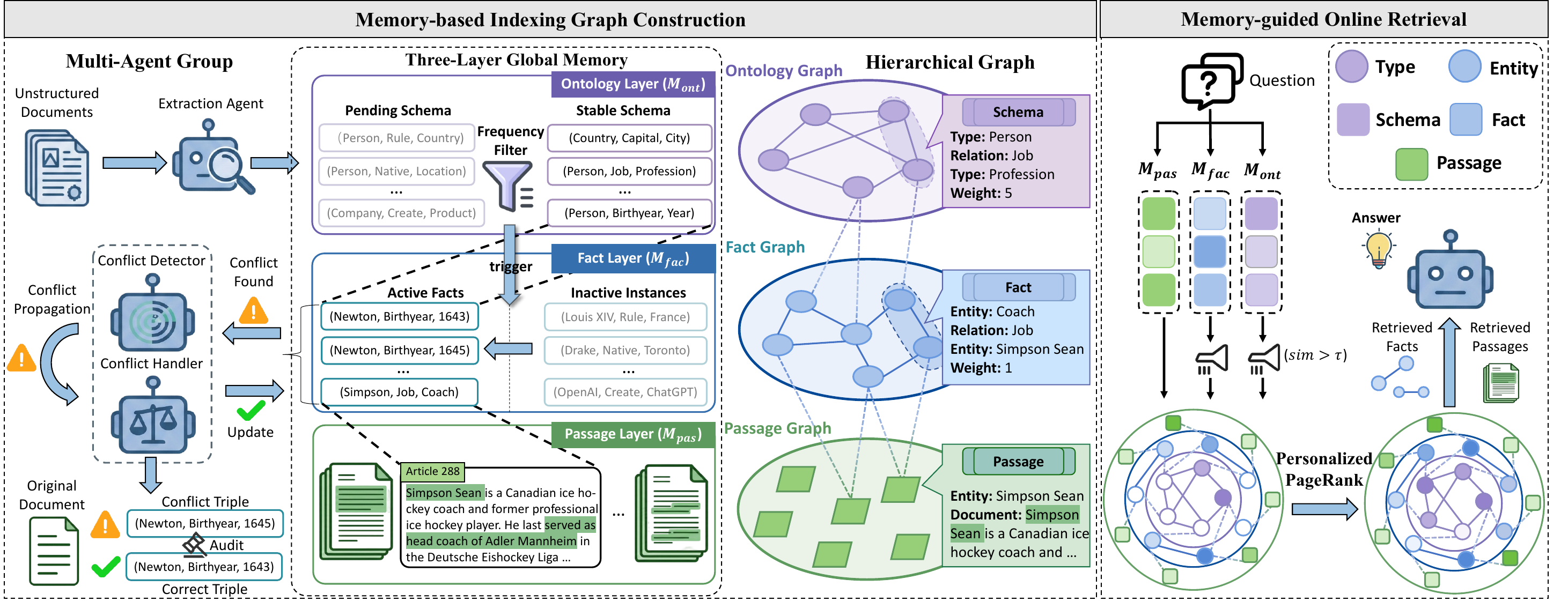}
    \caption{Overview of the MemGraphRAG framework with two phases: (i) Memory-Based Indexing Graph Construction, where Global Memory ($\mathcal{M}$) and the Knowledge Graph ($\mathcal{G}$) co-evolve via unified schema filtering, global adjudication, and memory-guided bridging; and (ii) Memory-Guided Online Retrieval, which leverages multi-layer memory filtering, structure-aware node initialization, and Personalized PageRank to identify globally relevant contexts for generation.}
    \label{fig:framework}
\end{figure*}
To overcome fragmented extraction and enable coherent graph evolution, we propose \textbf{MemGraphRAG}, a memory-based framework for constructing and maintaining high-quality knowledge graphs. Our key insight is that reliable graph construction requires not only structured storage, but also persistent coordination and correction across documents. As illustrated in Figure \ref{fig:framework}, it consists of two collaborative modules: Memory-based Graph Construction and Memory-guided Retrieval. We first introduce the foundational architecture, followed by the graph construction and retrieval pipelines.

\subsection{MemGraphRAG Architecture}\label{sec3.1}
MemGraphRAG consists of three core components: a \textit{Global Memory} that stores schemas, facts, and passages at different granularities and supports the construction of the \textit{Hierarchical Indexing Graph}, and a \textit{Multi-Agent Group} that interacts with memory to iteratively extract, detect, and resolve conflicts. Specifically:

\textbf{Global Memory ($\mathcal{M}$)} organizes the extracted knowledge into a three-tier hierarchy, including an \textit{Ontology Layer} ($\mathcal{M}_{ont}$) that stores schemas with extraction frequencies, a \textit{Fact Layer} ($\mathcal{M}_{fac}$) that maintains concrete facts, and a \textit{Passage Layer} ($\mathcal{M}_{pas}$) that preserves original text passages for evidence grounding. To strengthen cross-layer associations, we introduce a \textit{dense indexing mechanism} that enforces schema consistency and evidence traceability through two bidirectional interactions, where \textit{schema-instance alignment} links schemas with facts and \textit{fact-evidence grounding} connects facts with their supporting passages. (See more details in Appendix~\ref{appendix:indexing_mech}).

\textbf{Hierarchical Indexing Graph} ($\mathcal{G}$). It provides a unified representation spanning abstract schemas, concrete facts, and textual evidence. It consists of three interconnected graph views: (i) \textit{Semantic Ontology Graph} $\mathcal{G}_{ont}$, derived from $\mathcal{M}_{ont}$, which encodes schema-level type relations and structural constraints; (ii) \textit{Fact Graph} $\mathcal{G}_{fac}$, constructed from $\mathcal{M}_{fac}$, which represents instantiated entity-relation triples for multi-hop reasoning; and (iii) \textit{Source Evidence Graph} $\mathcal{G}_{pas}$, induced from $\mathcal{M}_{pas}$, which grounds facts in $\mathcal{G}_{fac}$ back to their supporting passages. This hierarchical design enables reasoning to traverse from abstract semantics to grounded evidence. More details are provided in Appendix~\ref{appendix:hierar_graph}.

\textbf{Multi-Agent Group} ($\mathcal{A}$). We introduce a group of agents $\mathcal{A}=\{A_{ext},A_{det},A_{res}\}$. Specifically: (i) the \textit{Extraction Agent} $A_{ext}$ extracts schemas, facts, and passages into $\mathcal{M}$ with evidence grounding; (ii) the \textit{Conflict Detection Agent} $A_{det}$ monitors $\mathcal{M}{fac}$ to detect redundancy, structural anomalies, and logical inconsistencies; and (iii) the \textit{Conflict Resolution Agent} $A{res}$ leverages schema constraints from $\mathcal{M}{ont}$ and historical evidence from $\mathcal{M}{pas}$ to resolve conflicts and maintain global consistency in $\mathcal{G}$. This design separates extraction, diagnosis, and correction for reliable graph construction.

\subsection{Memory-based Indexing Graph Construction} Traditional graph construction often processes document chunks in isolation, resulting in index fragmentation and noise accumulation. To address the critical limitations of \textit{Thematic Irrelevance}, \textit{Logical Inconsistency}, and \textit{Structural Fragmentation} identified in our pilot study, we reformulate knowledge graph construction as a dynamic co-evolution process between the Global Memory $\mathcal{M}$ and the Knowledge Graph $\mathcal{G}$. Distinct from static extraction pipelines, our approach adheres to three core principles designed to systematically resolve these issues: (i) \textbf{Thematic Denoising via Unified Schema Filtering}: Addressing \textit{Thematic Irrelevance}, we employ a unified schema to rigorously filter and manage extracted triples, ensuring that only thematically relevant knowledge is retained; (ii) \textbf{Consistency Maintenance via Global Adjudication}: To resolve \textit{Logical Inconsistency}, we utilize the global memory to assist agents in detecting and adjudicating semantic contradictions across disparate documents, thereby ensuring the logical unity of the graph; iii) \textbf{Structural Unification via Memory-Guided Bridging}: To overcome \textit{Structural Fragmentation}, we leverage the global memory to identify and merge equivalent entities across disconnected subgraphs. By connecting isolated local extractions and aligning them with the global ontology, we construct a cohesive and interconnected knowledge representation. Specifically, our graph construction procedure is described as follows:

\subsubsection{Thematic Denoising via Unified Schema Filtering}
Graph construction begins with the \textit{Extraction Agent} $A_{ext}$, which transforms each document chunk $c_i$ into structured memory entries. Rather than producing triples alone, $A_{ext}$ jointly constructs entries for all three layers of Global Memory$\mathcal{M}$ by generating candidate schemas, instantiated facts, and their supporting passages:
\begin{equation}\small
A_{ext}(c_i) \rightarrow \{ \mathcal{S}_{cand} \in \mathcal{M}_{ont}, \mathcal{T}_{cand} \in \mathcal{M}_{fac}, \mathcal{P}_{src} \in \mathcal{M}_{pas} \}.
\end{equation}
This design ensures that each extracted triple is strictly aligned with a schema and grounded in source evidence. To mitigate hallucination accumulation, newly generated schemas are initially treated as candidates and are promoted to stable schemas only when their empirical frequency exceeds a threshold:
\begin{equation}\small
\mathcal{M}_{ont}^{stable} = \{\, s \in \mathcal{M}_{ont} \mid \mathrm{Freq}(s) \ge \tau \,\}.
\end{equation}
Accordingly, only facts aligned with stable schemas are activated for downstream graph construction and reasoning. Detailed extraction procedures are provided in Appendix~\ref{appendix:dinoising}.

\subsubsection{Consistency Maintenance via Global Adjudication}
During evolutionary extraction, newly activated triples may introduce redundancy or semantic conflicts with existing facts. To ensure the long-term consistency of the \textit{Fact Layer} $\mathcal{M}_{fac}$, We deploy a decoupled diagnosis and correction loop, where the \textit{Conflict Detection Agent} ($A_{det}$) and the \textit{Conflict Resolution Agent} ($A_{res}$) collaborate to continuously maintain memory integrity. Specifically, when a new triple $t_{new} \in \mathcal{M}_{fac}$ becomes active, $A_{det}$ asynchronously scans existing facts and identifies a conflict set $\mathcal{F}_{conf}$ based on semantic similarity and ontology-level structural constraints:
\begin{equation}\small
\mathcal{F}_{conf} = \{\, t' \in \mathcal{M}_{fac} \mid \mathrm{Sim}(t_{new}, t') > \delta \;\lor\; \mathrm{Match}(t_{new}, t') \,\}.
\end{equation}
If $\mathcal{F}_{conf}$ is non-empty, $A_{res}$ is triggered to resolve the detected inconsistencies. Rather than generating corrections heuristically, $A_{res}$ leverages \textit{fact-evidence grounding} to retrieve the provenance passages from $\mathcal{M}_{pas}$ and adjudicates conflicts by comparing the corresponding textual evidence. This evidence-driven resolution enables reliable corrective actions such as filtering invalid facts, merging redundant triples, and resolving temporal or granularity inconsistencies, thereby ensuring that $\mathcal{M}_{fac}$ remains globally coherent throughout the graph construction process.

\subsubsection{Structural Unification via Memory-Guided Bridging}
In the final phase, we project the refined Global Memory $\mathcal{M}$ into the \textit{Hierarchical Indexing Graph} $\mathcal{G}$ by constructing three interconnected graph views. Specifically, we build the \textit{Semantic Ontology Graph} $\mathcal{G}_{ont}$ directly from $\mathcal{M}_{ont}$, where nodes and edges encode schema-level types and their valid relations, serving as the logical backbone of the overall structure. We then construct the \textit{Fact Graph} $\mathcal{G}_{fac}$ from $\mathcal{M}_{fac}$, where entities form nodes and instantiated triples form edges, enabling multi-hop reasoning over concrete facts. To improve connectivity and reduce fragmentation, we further augment $\mathcal{G}_{fac}$ by introducing additional bridging edges, including type-based connections derived from shared stable schema types in $\mathcal{G}_{ont}$ and similarity-based connections between entities with high embedding similarity. Finally, we induce the \textit{Source Evidence Graph} $\mathcal{G}_{pas}$ from $\mathcal{M}_{pas}$, which links facts and entities in $\mathcal{G}_{fac}$ back to their originating passages, ensuring that every reasoning path remains traceable to grounded textual evidence.

\definecolor{ourrowbg}{RGB}{235, 242, 255} 

\definecolor{gainlow}{RGB}{235, 250, 235}
\definecolor{gainmid}{RGB}{200, 245, 200}
\definecolor{gainhigh}{RGB}{160, 230, 160}

\newcommand{\cghigh}[1]{\cellcolor{gainhigh}\textcolor{black}{+#1}}
\newcommand{\cgmid}[1]{\cellcolor{gainmid}\textcolor{black}{+#1}}
\newcommand{\cglow}[1]{\cellcolor{gainlow}\textcolor{black}{+#1}}

\begin{table*}[t]
    \caption{Generation performance of different GraphRAG methods. The best result is \textbf{bold}, and the second is \underline{underline}. The column \textbf{$\Delta$} indicates the performance gain of our \textbf{MemGraphRAG} (59.25) compared to each baseline. Background colors in \textcolor[rgb]{0.56, 0.93, 0.56}{$\Delta$ columns} represent the magnitude of improvement (Darker green = larger gap).} 
    \label{tab:main_results}
    \small
    \centering
    \vspace{-3mm}
    \scalebox{1}{ 
    \begin{tabular}{lcccccccccc} 
    \toprule
    \multirow{2}{*}{\textbf{Method}}
        &\multicolumn{2}{c}{\textbf{HotpotQA}}
        & \multicolumn{2}{c}{\textbf{2WikiMultiHopQA}} 
        &\multicolumn{2}{c}{\textbf{MuSiQue}}
        & \textbf{G-Medical}
        & \textbf{G-Novel}
        & \multicolumn{2}{c}{\textbf{Overall}}\\ 
        \cmidrule(lr){2-3} \cmidrule(lr){4-5} \cmidrule(lr){6-7} \cmidrule(lr){8-8} \cmidrule(lr){9-9} \cmidrule(lr){10-11}
        &Str-Acc. &LLM-Acc. &Str-Acc. &LLM-Acc. &Str-Acc. &LLM-Acc. &LLM-Acc. &LLM-Acc. & Avg. & $\Delta$ \\
    \midrule 
    \multicolumn{11}{c}{\textbf{\textit{Direct Zero-shot LLM Inference}}} \\
    \midrule
        Llama3-8B   & 30.80 & 28.20 & 34.00 & 16.00 & 6.70  & 7.40  & 26.43 & 15.20 & 20.59 & \cghigh{38.66} \\
        Llama3-13B  & 24.90 & 17.00 & 22.30 & 9.50  & 4.20  & 5.00  & 28.28 & 19.30 & 16.31 & \cghigh{42.94} \\
        GPT-3.5-Turbo & 32.70 & 42.50 & 28.30 & 31.00 & 10.10 & 21.40 & 45.82 & 29.41 & 30.15 & \cghigh{29.10} \\
        GPT-4o-mini & 38.10 & 39.70 & 36.00 & 31.30 & 14.10 & 15.20 & 42.13 & 31.42 & 30.99 & \cghigh{28.26} \\
    \midrule
    \multicolumn{11}{c}{\textbf{\textit{Vanilla Retrieval-Augmented-Generation}}} \\
    \midrule
        Retrieval (Top-1)   & 48.80 & 50.40 & 38.10 & 34.00 & 19.90 & 23.80 & 50.90 & 43.94 & 38.73 & \cghigh{20.52} \\
        Retrieval (Top-3)   & 55.40 & 58.90 & 46.50 & 41.80 & 26.60 & 28.50 & 55.16 & 46.06 & 44.87 & \cgmid{14.38} \\
        Retrieval (Top-5)   & 58.50 & 60.30 & 49.80 & 45.40 & 28.30 & 32.00 & 61.07 & 48.35 & 47.97 & \cgmid{11.28} \\
    \midrule
    \multicolumn{11}{c}{\textbf{\textit{Graph-based Retrieval-Augmented-Generation Methods}}} \\
    \midrule
        KGP~\cite{wang2024knowledge}         & 62.70 & 62.10 & 33.10 & 32.70 & 28.40 & 32.50 & 56.29 & 49.01 & 44.60 & \cgmid{14.65} \\
        G-retriever~\cite{he2024g}           & 44.00 & 41.80 & 47.80 & 29.70 & 16.20 & 17.60 & 52.40 & 45.90 & 36.93 & \cghigh{22.32} \\
        RAPTOR~\cite{sarthi2024raptor}       & 57.00 & 61.00 & 51.70 & 43.60 & 24.70 & 28.90 & 57.88 & 44.24 & 46.13 & \cgmid{13.12} \\
        MS-GraphRAG~\cite{edge2024local}     & 51.60 & 43.50 & 47.30 & 38.60 & 20.60 & 23.70 & 55.67 & 50.43 & 41.43 & \cghigh{17.82} \\
        LazyGraphRAG~\cite{Lazygraphrag}     & 52.70 & 43.80 & 46.80 & 37.90 & 21.50 & 24.80 & 56.63 & 51.56 & 41.96 & \cghigh{17.29} \\
        LightRAG~\cite{guo2024lightrag}      & 61.40 & 62.00 & 56.90 & 40.50 & 28.60 & 30.50 & 56.42 & 46.09 & 47.80 & \cgmid{11.45} \\
        HippoRAG~\cite{hipporag}             & 58.40 & 61.40 & 67.50 & 61.30 & 30.40 & 26.00 & 57.06 & 45.77 & 50.98 & \cgmid{8.27} \\
        HippoRAG2~\cite{gutiérrez2025hipporag2} & 65.20 & 67.20 & 64.20 & 57.90 & 32.20 & \textbf{38.30} & 64.85 & \underline{56.48} & 55.79 & \cglow{3.46} \\
        E$^2$GraphRAG~\cite{zhao20252graphrag}  & 63.10 & 65.70 & 57.20 & 40.90 & 26.10 & 29.00 & 60.24 & 54.28 & 49.57 & \cgmid{9.68} \\
        GFM-RAG~\cite{luo2025gfm}               & 64.10 & \underline{67.70} & 69.10 & 61.10 & 32.50 & 36.10 & 58.19 & 53.39 & 55.27 & \cglow{3.98} \\
        LogicRAG~\cite{chen2025logicrag}        & 55.80 & 65.60 & 64.80 & 63.40 & 30.10 & 34.60 & 56.75 & 49.84 & 52.61 & \cgmid{6.64} \\
        LinearRAG~\cite{zhuang2025linearrag}    & \underline{65.30} & 67.30 & \underline{70.20} & \underline{65.70} & \underline{33.20} & 37.20 & \underline{65.70} & 52.57 & \underline{57.15} & \cglow{2.10} \\
    \midrule
    \rowcolor{ourrowbg}
        \textbf{MemGraphRAG (Ours)} & \textbf{67.20} & \textbf{71.60} & \textbf{70.30} & \textbf{69.80} & \textbf{34.40} & \underline{37.90} & \textbf{68.40} & \textbf{57.41} & \textbf{59.25} & \textbf{--} \\
    \bottomrule
    \end{tabular}}
\end{table*}

\subsection{Memory-guided Online Retrieval}
Building upon the Global Hierarchical Graph $\mathcal{G}$ and Global Memory $\mathcal{M}$, we perform memory-guided retrieval and reasoning in three stages: (i) \textbf{Multi-Layer Memory Retrieval}, which retrieves candidate schemas, facts, and passages from $\mathcal{M}_{ont}$, $\mathcal{M}_{fac}$, and $\mathcal{M}_{pas}$; (ii) \textbf{Structure-Aware Node Initialization}, which maps the retrieved evidence to initial node weights based on semantic relevance and structural signals; and (iii) \textbf{Graph Propagation}, which runs Personalized PageRank (PPR) over the heterogeneous graph to rank globally important nodes and passages for  LLM generation.


\subsubsection{Multi-Layer Memory Filtering}
The retrieval phase initiates by querying the three distinct layers of the Global Memory$\mathcal{M}$ in parallel. Given a user query $\mathbf{q}$, we retrieve top-$K$ candidates from $\mathcal{M}$ in parallel, including schemas from $\mathcal{M}_{ont}$, facts from $\mathcal{M}_{fac}$, and passages from $\mathcal{M}_{pas}$. To reduce noise before graph reasoning, we retain only schemas and facts whose semantic similarity satisfies $\mathrm{Sim}(\mathbf{q}, \mathbf{x}) > \tau$. This filtering ensures that subsequent node initialization is seeded with high-confidence structural evidence. If no valid structural candidates remain (i.e., $\mathcal{S}_{ret} \cup \mathcal{F}_{ret} = \emptyset$), we fall back to standard RAG retrieval by directly selecting passages from $\mathcal{M}_{pas}$ based on query similarity.

\subsubsection{Structure-Aware Node Initialization} 
To seed graph propagation with query-specific context, we project the retrieved evidence onto the heterogeneous graph by defining an initial reset probability distribution $P_{init}(v)$ for each node $v \in \mathcal{G}$. This distribution assigns the starting importance of nodes before propagation. We then initialize $P_{init}(v)$ along three complementary dimensions, as detailed below.

\noindent\textbf{Entity Node Initialization via Facts:} To ensure that graph propagation originates from grounded evidence, we initialize each entity node $e$ based on the relevance of its associated retrieved facts. Specifically, its initial weight is defined as the mean similarity over all query-relevant facts containing $e$:
\begin{equation}\small
    P_{init}(e) = \frac{1}{|\mathcal{F}_e|} \sum_{f \in \mathcal{F}_e} \mathrm{Sim}(\mathbf{q}, \mathbf{f}),
\end{equation}
where $\mathcal{F}_e \subseteq \mathcal{F}_{ret}$ denotes the subset of retrieved facts that contain entity $e$. If $\mathcal{F}_e=\emptyset$, we set $P_{init}(e)=0$.

\noindent\textbf{Type Node Initialization via Schemas:} We further initialize type nodes $t \in \mathcal{G}_{\text{schema}}$ based on the retrieved schemas from $\mathcal{M}_{\text{ont}}$ to avoid introducing irrelevant semantics. A critical challenge is that type nodes often exhibit exceptionally large degrees (e.g., a generic ``Person'' node connected to thousands of entities). Directly activating such high-degree nodes would spread importance across overly many nodes, introducing significant noise. To address this issue, we introduce a structural regularization term that combines semantic relevance with a log-degree penalty:

\begin{equation}\small
    P_{init}(t) = \underbrace{\left( \frac{1}{|\mathcal{S}_t|} \sum_{s \in \mathcal{S}_t} \text{Sim}(\mathbf{q}, \mathbf{s}) \right)}_{\text{Schema Relevance}} \times \underbrace{\frac{1}{\log(\text{deg}(t) + 1)}}_{\text{Hub Suppression}}
\end{equation}

where $\mathcal{S}_t$ denotes the retrieved schemas associated with $t$. This design incorporates schema-level relevance while preventing overly generic types from dominating propagation.

\noindent\textbf{Passage Node Initialization via Information Density:} Finally, we initialize the Passage Nodes ($p \in G_{pas}$) by combining semantic relevance with an information density prior:
\begin{equation}\small
    P_{init}(p) = \text{Sim}(\mathbf{q}, \mathbf{d}_p) \times \alpha \times \underbrace{\sigma\left( \frac{\sum_{e \in \mathcal{E}_p} \text{IDF}(e)}{\log(|\mathcal{E}_p| + 1)} \right)}_{\text{Information Density Term}}
\end{equation}
This scoring function combines semantic alignment $\mathrm{Sim}(\mathbf{q}, \mathbf{d}_p)$, a dampening factor $\alpha$ (set to 0.05) to prevent passage nodes from dominating propagation, and an \textit{Information Density Term} that favors passages containing rare and informative entities by aggregating their IDF scores with log-normalization. Detailed initialization procedures are provided in Appendix~\ref{appedix:retrieval_details}.

\subsubsection{Personalized PageRank}
After initialization, we run Personalized PageRank (PPR) on the heterogeneous graph to propagate query-specific importance. Starting from the normalized distribution $\mathbf{v}^{(0)}$, the iteration is defined as $\mathbf{v}^{(k+1)} = (1 - \lambda)\mathbf{W}\mathbf{v}^{(k)} + \lambda \mathbf{v}^{(0)}$, where $\mathbf{W}$ denotes the transition matrix and $\lambda$ is the damping factor. We set $\lambda=0.5$ to limit propagation within a local neighborhood and reduce semantic drift. After convergence, we select the top-$K$ passages and top-$M$ entities ranked by $\mathbf{v}^{(\infty)}$ for LLM inference.

\begin{table*}[h]
\caption{Retrieval performance of different GraphRAG methods on G-Bench(Medical).} 
\label{tab:retrieval}
\vspace{-2mm}
\centering
\setlength{\tabcolsep}{5.pt} 
\begin{tabular}{lccccccccc}
\toprule
\multirow{2}{*}{\textbf{Method}} & \multicolumn{2}{c}{\textbf{Fact Retrieval}} & \multicolumn{2}{c}{\textbf{Complex Reasoning}} & \multicolumn{2}{c}{\textbf{Contextual}} & \multicolumn{2}{c}{\textbf{Creative Gen}} & \multirow{2}{*}{\textbf{Retrieval Time}} \\
\cmidrule(lr){2-3} \cmidrule(lr){4-5} \cmidrule(lr){6-7} \cmidrule(lr){8-9}
 & Recall & Relevance & Recall & Relevance & Recall & Relevance & Recall & Relevance & \\
\midrule
RAPTOR~\cite{sarthi2024raptor}          & 85.40  & 69.38 & \underline{89.70}  & 53.20  & 88.86 & 58.73 & 72.70  & 52.71 & 0.171  \\
Lazy-GraphRAG~\cite{Lazygraphrag}       & 74.29 & 19.90  & 78.65 & 17.50  & 78.72 & 21.35 & 83.41 & 15.09 & 9.835  \\
LightRAG~\cite{guo2024lightrag}         & 80.32 & 41.27 & 82.91 & 42.79 & 85.71 & 43.11 & 81.34 & 45.17 & 11.052 \\
HippoRAG~\cite{hipporag}                & 87.25 & 52.44 & 83.80  & 42.19 & 83.46 & 49.13 & 81.66 & 45.03 & 1.586  \\
HippoRAG2~\cite{gutiérrez2025hipporag2} & 78.70  & \underline{87.96} & 77.00    & 80.94 & 77.40  & 86.85 & 61.12 & \underline{78.64} & 2.157  \\
GFM-RAG~\cite{luo2025gfm}               & \textbf{90.08} & 57.90  & 85.03 & 33.06 & 78.62 & 40.14 & 83.51 & 22.87 & 1.375  \\
LinearRAG~\cite{zhuang2025linearrag}    & 88.86 & 86.09 & 87.03 & \underline{81.58} & \underline{89.13} & \textbf{87.89} & \underline{89.08} & 72.74 & \underline{0.123}  \\
\midrule
\textbf{MemGraphRAG(ours)}                           & \underline{89.56} & \textbf{88.53} & \textbf{90.42} & \textbf{82.64} & \textbf{89.57} & \underline{86.91} & \textbf{89.86} & \textbf{79.12} & \textbf{0.061}  \\
\bottomrule
\end{tabular}%
\end{table*}

\section{Experiments}
In this section, our aim is to answer the following questions: \textbf{Q1} (Generation Accuracy): How does MemGraphRAG perform compared to state-of-the-art GraphRAG methods in terms of generation performance? \textbf{Q2} (Retrieval Analysis): How does our retrieval method compare to other frameworks in terms of performance and efficiency?\textbf{Q3} (Graph Adaptability Analysis): Can the graph constructed by MemGraphRAG generalize to other GraphRAG methods? \textbf{Q4} (Ablation Study): What contribution does each component of MemGraphRAG make to the overall performance? (Note that additional experiments and case studies are provided in Appendix~\ref{app:more_exp}.)
\subsection{Experimental Setting}

\textbf{Datasets.} We first evaluate the effectiveness of MemGraphRAG on three widely-used multi-hop QA datasets, including HotpotQA~\cite{yang2018hotpotqa}, 2WikiMultiHopQA (2Wiki)~\cite{2wikimqa}, MuSiQue~\cite{trivedi2022MuSiQue}. We follow the settings used in \cite{hipporag,gutiérrez2025hipporag2} for a fair comparison, choosing 1,000 questions from each validation set. We also test our approach on G-Bench(Medical) and G-bench(Novel)~\cite{xiang2025use} to evaluate MemGraphRAG on complex reasoning across medical, novel knowledge. More details about datasets can be found in Appendix~\ref{app:dataset}.

\noindent\textbf{Baselines.} We categorize all baselines into three groups: (i) Zero-shot LLM Inference: We evaluate several foundational models including LLaMA3 (8B) and LLaMA3 (13B) ~\cite{dubey2024llama3}, as well as GPT-3.5-turbo and GPT-4o-mini~\cite{openai2023gpt4}. (ii) We deploy Vanilla RAG across multiple retrieval configurations (retrieving 1, 3, or 5 top passages). (iii) State-of-the-art GraphRAG Systems: We compare against leading GraphRAG implementations, including KGP ~\cite{wang2024knowledge}, G-retriever~\cite{he2024g}, LightRAG~\cite{guo2024lightrag}, RAPTOR ~\cite{sarthi2024raptor}, MS-GraphRAG\cite{edge2024local}, HippoRAG ~\cite{hipporag,gutiérrez2025hipporag2}, GFM-RAG~\cite{luo2025gfm}, LazyGRAG\cite{Lazygraphrag}, E$^2$GraphRAG~\cite{zhao20252graphrag},   LogicRAG\cite{chen2025logicrag} and LinearRAG\cite{zhuang2025linearrag}. 

\noindent\textbf{Evaluation Metrics.} We evaluate our method using four metrics across two categories. For QA performance, following existing work\cite{chen2025logicrag, zhuang2025linearrag}, we use: 1) String-based accuracy (Str-Acc.), which computes whether the gold answer is included in the generated answer after normalizing them to lowercase words, and 2) LLM-based accuracy (LLM-Acc.), which lets an LLM decide whether the generated answer correctly matches the gold answer. For GraphRAG-bench, since golden answers consist of lengthy descriptive statements, we only evaluate using LLM-ACC. For retrieval quality assessment, we adopt metrics from GraphRAG-Bench~\cite{xiang2025use}: 1) \textit{Context Relevance}, which measures semantic alignment between questions and retrieved passages, and 2) \textit{Evidence Recall}, which evaluates whether the retrieved contents contain all the necessary information that used for generating the correct answer.

\noindent\textbf{Implementation Details.} For consistency, all methods use the same embedding model (\textit{i.e.}, NV-Embed-v2~\cite{moreira2024nv}). We set $k$=5 for top-$k$ retrieval in all methods. For both offline indexing (graph construction) and online generation, we adopt GPT-4o-mini as the default LLM (additional open-source LLM results are reported in Appendix~\ref{app:more_exp}). For evaluation, we use GPT-4o-mini to compute the LLM-Acc metric. To ensure reproducibility, we set the inference temperature to 0 for all LLM calls.

\subsection{Generation Accuracy (Q1)}
To address Q1, we conduct a comprehensive evaluation of generation performance by comparing various baseline methods with MemGraphRAG across four benchmark datasets. The detailed experimental results are presented in Table \ref{tab:main_results}. Based on our analysis, we derive the following key observations.

\textbf{RAG system significantly enhances the LLM generation performance.} Direct inference (without retrieval) yields the lowest scores across all benchmarks. For instance, GPT-4o-mini achieves a mere 14.65\% average accuracy on MuSiQue in a zero-retrieval setting. Integrating retrieved contexts via Vanilla RAG (top-5) doubles this performance to 30.15\%. This confirms that retrieval augmentation is essential for knowledge-intensive tasks.

\textbf{Graph-based retrieval is more effective for multi-hop reasoning.} While increasing the retrieval count ($k$) improves Vanilla RAG, the performance gains quickly plateau. This limitation stems from Vanilla RAG’s reliance on surface-level keyword matching, which often overlooks the logical bridges required for multi-hop reasoning. In contrast, GraphRAG methods explicitly capture structural dependencies and consistently, and often deliver stronger results. Notably, HippoRAG 2 emerges as a competitive baseline, achieving 38.30\% and 56.48\% LLM-based accuracy on MuSiQue and G-novel, respectively.

\textbf{MemGraphRAG consistently surpasses existing GraphRAG baselines.} While exiting GraphRAGs attempt to align semantics through graph structures, they are often sensitive to noise and low-quality indexing introduced by solated chunk-
level extraction. In contrast, MemGraphRAG mitigates these issues by providing more reliable indexing and retrieval, achieves the best results across all datasets.  It reaches 59.25\% average accuracy, yielding a 2.10\% absolute gain over the strongest baseline.

\definecolor{rowblue}{RGB}{236, 246, 255} 
\definecolor{rowpurple}{RGB}{245, 240, 255}
\definecolor{textgreen}{RGB}{0, 140, 60} 
\definecolor{deepgreen}{RGB}{0, 100, 30}

\begin{table*}[t]
\centering
\caption{Adaptability Analysis: MemGraphRAG as a universal graph constructor across different frameworks. The \colorbox{rowblue}{blue rows} indicate experiments using MemGraphRAG's constructed graph, while the \colorbox{rowpurple}{purple row} represents our full framework. The rightmost column shows the performance gain.} 
\label{tab:graph-comparison}
\vspace{-3mm}
\setlength{\tabcolsep}{6.1pt}
\begin{tabular}{ll|cccccc>{\bfseries}c}
\toprule
\textbf{GraphConstructor} & \textbf{Retriever} 
& \textbf{HotpotQA} 
& \textbf{2Wiki} 
& \textbf{MuSiQue} 
& \textbf{G-Medical} 
& \textbf{G-Novel} 
& \textbf{Average} 
& \textcolor{black}{$\Delta$} \\
\midrule

HippoRAG~\cite{hipporag} & HippoRAG~\cite{hipporag} 
& 59.90 & 64.40 & 28.20 & 57.06 & 45.77 & 51.07 & \textcolor{textgreen}{+8.61} \\
\rowcolor{rowblue}
MemGraphRAG & HippoRAG~\cite{hipporag} 
& 60.65 & \underline{65.25} & 29.00 & 57.75 & 46.24 & 51.78 & \textcolor{textgreen}{+7.90} \\

\addlinespace[3pt]

HippoRAG2~\cite{gutiérrez2025hipporag2} & HippoRAG2~\cite{gutiérrez2025hipporag2} 
& \underline{66.20} & 61.05 & 35.25 & 64.85 & \underline{56.48} & 56.77 & \textcolor{textgreen}{+2.91} \\
\rowcolor{rowblue}
MemGraphRAG & HippoRAG2~\cite{gutiérrez2025hipporag2} 
& 66.00 & 61.20 & \underline{35.40} & \underline{65.42} & \textbf{56.76} & \underline{56.96} & \textcolor{textgreen}{+2.72} \\

\addlinespace[3pt]

MS-GraphRAG~\cite{edge2024local} & MS-GraphRAG~\cite{edge2024local} 
& 47.55 & 42.95 & 22.15 & 55.67 & 50.43 & 43.75 & \textcolor{textgreen}{+15.93} \\
\rowcolor{rowblue}
MemGraphRAG & MS-GraphRAG~\cite{edge2024local} 
& 48.00 & 43.20 & 22.45 & 56.53 & 50.88 & 44.21 & \textcolor{textgreen}{+15.47} \\

\addlinespace[3pt]

LazyGraphRAG~\cite{Lazygraphrag} & LazyGraphRAG~\cite{Lazygraphrag} 
& 48.25 & 42.35 & 23.15 & 56.63 & 51.56 & 44.39 & \textcolor{textgreen}{+15.29} \\
\rowcolor{rowblue}
MemGraphRAG & LazyGraphRAG~\cite{Lazygraphrag} 
& 48.75 & 42.55 & 23.50 & 57.98 & 52.06 & 44.97 & \textcolor{textgreen}{+14.71} \\

\midrule 

\rowcolor{rowpurple}
MemGraphRAG & MemGraphRAG 
& \textbf{69.40}  & \textbf{70.05} & \textbf{36.15} & \textbf{68.40} & 54.41 & \textbf{59.68} & - \\

\bottomrule
\end{tabular}
\vspace{-3mm}
\end{table*}

\subsection{Retrieval Analysis (Q2)}

To evaluate the retrieval performance of MemGraphRAG, we conducted tests across four distinct task levels on the GraphRAG-Bench. We utilized Recall and Relevance as metrics to assess the GraphRAG's capacity for retrieving both comprehensive and precise information. Additionally, to assess practical deployment feasibility, we recorded the average retrieval time (in seconds) across all queries. The experimental results are presented in Table~\ref{tab:retrieval}.

\textbf{MemGraphRAG achieves consistently strong retrieval performance, balancing high recall with high relevance.} MemGraphRAG consistently ranks at the top in \textit{Complex Reasoning} tasks (Recall: 90.42, Relevance: 82.64) and \textit{Fact Retrieval} tasks, significantly outperforming baselines such as HippoRAG2 and LightRAG. These results indicate that our approach effectively filters noise and invalid entity relationships, enabling the system to precisely pinpoint entities and relations relevant to the query. Unlike methods that sacrifice precision for coverage, MemGraphRAG maintains superior relevance while capturing broad context, thereby validating the effectiveness of our \textit{Global Adjudication} mechanism for consistency maintenance in constructing high-quality graphs.

\textbf{MemGraphRAG achieves the lowest retrieval latency, showing superior online inference efficiency.} MemGraphRAG requires an average of only 0.061 seconds per retrieval, which is significantly faster than LightRAG (11.052s) and HippoRAG (1.586s). This efficiency is attributed to our lightweight retrieval process, which relies on efficient Personalized PageRank (PPR) rather than computationally expensive real-time LLM filtering or iterative reasoning loops. Consequently, MemGraphRAG delivers high-precision complex reasoning while maintaining low latency in practice. 

\subsection{Indexing Graph Adaptability Analysis (Q3)}
To evaluate whether our constructed index graph can seamlessly adapt to different GraphRAG frameworks, we conducted a transferability experiment. Our pilot study previously identified that existing graph construction methods suffer from critical deficiencies, including Thematic Irrelevance, Logical Inconsistency, and Structural Fragmentation. Consequently, we assess whether the structural unification enabled by MemGraphRAG through Memory-Guided Bridging can mitigate these issues for other frameworks. Specifically, we replaced the native graph construction modules of HippoRAG, HippoRAG2, MS-GraphRAG, and LazyGraphRAG with the graph constructed by MemGraphRAG, while retaining their original downstream retrieval and reasoning mechanisms. The comparative experimental results are presented in Table \ref{tab:graph-comparison}.

\begin{figure}[t] 
    \centering  
    \includegraphics[width=\linewidth]{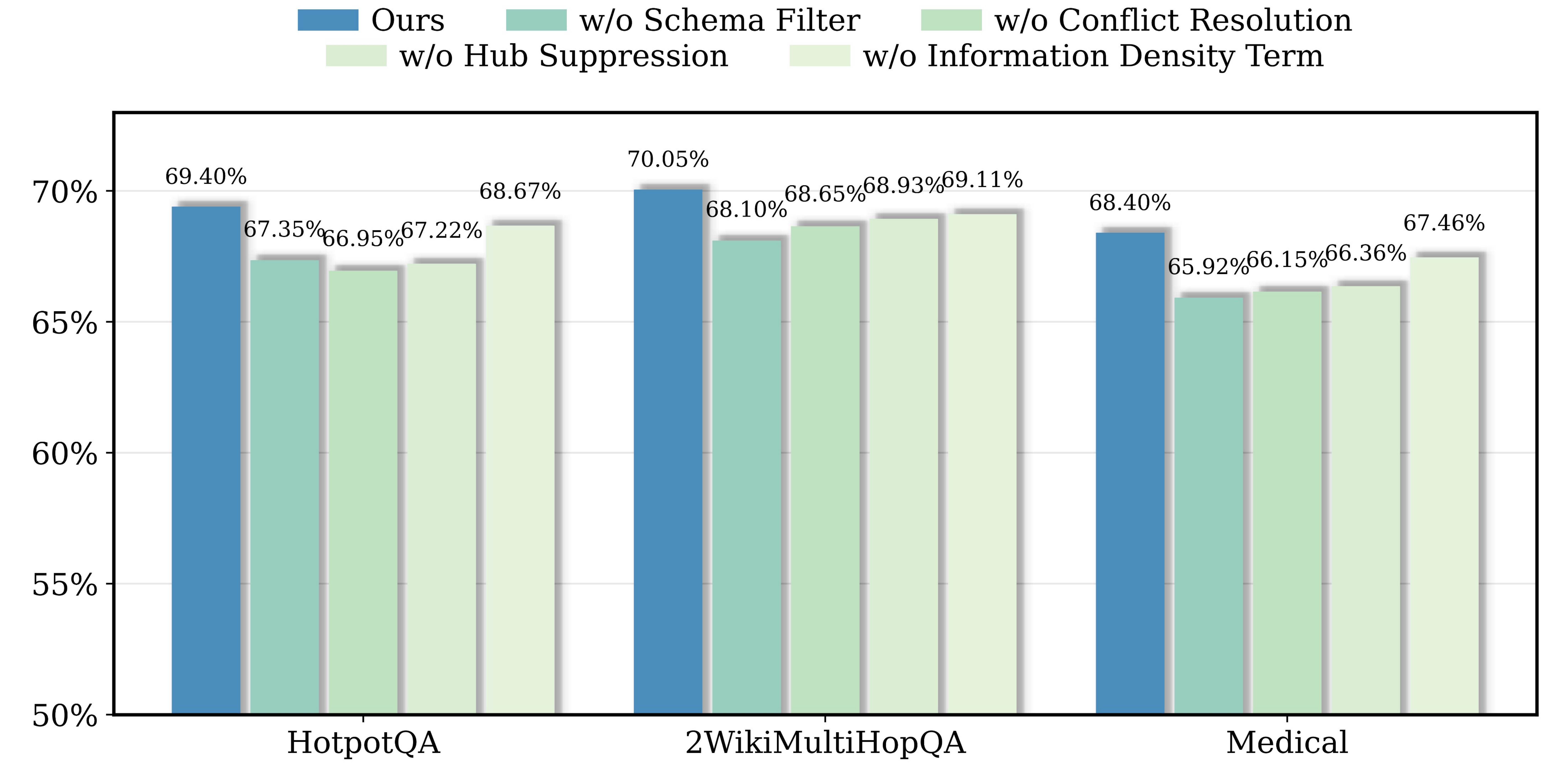} 
     \vspace{-5mm}
    \caption{Ablation study of MemGraphRAG on three datasets.}
    \label{fig:ablation}
    \vspace{-5mm}
\end{figure}
\textbf{MemGraphRAG consistently improves retrieval performance across all evaluated GraphRAG frameworks, serving as a universal high-quality graph constructor.} As shown in the results, replacing the original graphs with MemGraphRAG-constructed graphs leads to consistent improvements for all baseline retrievers across all datasets. For instance, the average performance of HippoRAG increases from 51.07 to 51.78, and MS-GraphRAG improves from 43.75 to 44.21. This consistent improvement shows that our memory-driven global construction mechanism effectively mitigates structural fragmentation and logical inconsistencies. By producing a more cohesive and thematically aligned knowledge structure, MemGraphRAG substantially strengthens the effectiveness of existing retrievers, demonstrating its robustness as a foundational indexing solution.

\subsection{Ablation Study (Q4)}

To verify the contribution of each module in MemGraphRAG, we conduct an ablation study on HotpotQA, 2WikiMultiHopQA, and G-Medical. We compare the full model with four variants that remove the \textit{Schema Filter}, \textit{Conflict Resolution}, \textit{Hub Suppression}, and the \textit{Information Density Term}, respectively. As shown in Figure~\ref{fig:ablation}, MemGraphRAG consistently achieves the best performance across all datasets (e.g., 69.40\% on HotpotQA), which indicates that these memory-driven graph construction and initialization mechanisms are jointly crucial for building a robust knowledge graph.

\textbf{w/o Schema Filter:} Removing \textit{Unified Schema Filtering} causes a clear degradation, especially on 2WikiMultiHopQA and G-Medical (68.10\% and 65.92\%). Without the frequency-based stability constraint ($\mathrm{Freq}(s)\ge\tau$), low-frequency and off-topic schemas are retained, introducing noisy triples that weaken semantic focus.

\textbf{w/o Conflict Resolution:} Excluding \textit{Global Adjudication} leads to the largest drop on HotpotQA (66.95\%). Without conflict detection and resolution, the fact layer accumulates contradictory or redundant triples, which disrupts multi-hop reasoning chains and increases the chance of retrieving inconsistent evidence.

\textbf{w/o Hub Suppression:} Removing \textit{Hub Suppression} reduces accuracy (67.22\% on HotpotQA). Without degree-based regularization, generic high-degree nodes dominate propagation, causing semantic drift toward irrelevant subgraphs.

\textbf{w/o Information Density Term:} Dropping the \textit{Information Density Term} yields a smaller but consistent decline (68.67\% on HotpotQA). Without IDF-style weighting, passage initialization cannot prioritize discriminative evidence, weakening the model’s ability to anchor reasoning on informative documents.

\section{Conclusion}
In this paper, we propose MemGraphRAG, a novel GraphRAG framework that integrates a global memory mechanism into the knowledge graph construction process. By leveraging a shared hierarchical memory structure, our multi-agent system collaboratively maintains a global perspective throughout both the extraction and retrieval phases. This paradigm effectively overcomes key limitations of traditional GraphRAG approaches that rely on isolated local extraction. It systematically mitigates thematic irrelevance, logical inconsistency, and structural fragmentation, thereby enabling a globally consistent indexing graph. Extensive experiments demonstrate that MemGraphRAG consistently outperforms state-of-the-art baselines in terms of graph quality, retrieval precision, and generation accuracy, providing a robust solution for deploying reliable RAG systems in complex real-world scenarios.

\section*{Limitation}
While MemGraphRAG demonstrates strong robustness in processing large-scale textual corpora and constructing globally consistent knowledge graphs, its current design is limited to unimodal textual inputs. However, real-world knowledge repositories are inherently multimodal, containing heterogeneous formats such as statistical charts, technical diagrams, document layouts, and embedded images in academic papers or financial reports. Currently, our framework requires non-textual elements to be transcribed or described in text before processing, which may lead to the loss of critical visual semantics and spatial relationships. For example, quantitative trends in line charts or complex structures in scientific diagrams often contain dense information that textual descriptions cannot fully capture, potentially causing information loss during indexing. Extending the \textit{Global Hierarchical Graph} to incorporate multimodal nodes (e.g., embedding visual patches into the \textit{Fact Layer} $\mathcal{M}_{fac}$ or the \textit{Passage Layer} $\mathcal{M}_{pas}$) is a promising direction for future work. Such an extension could enable cross-modal reasoning, allowing the multi-agent system to verify textual claims against visual evidence and further improve the versatility of MemGraphRAG.

\begin{acks}
The project was supported by 
Natural Science Foundation of Fujian Province of China (No. 2024J011001)
and
the Public Technology Service Platform Project of Xiamen (No.3502Z20231043).
We also thank the reviewers for their insightful comments.
\end{acks}
\balance 
\bibliography{ref}
\bibliographystyle{ACM-Reference-Format}


\appendix

\section{Additional Experiments}\label{app:more_exp}

\subsection{Ablation on Backbone LLMs}
To further evaluate the universality and robustness of MemGraphRAG, we conducted experiments utilizing the stronger \texttt{llama-3-70b- instruct} as the underlying backbone model. We compared our method against a comprehensive suite of baselines, ranging from non-structured methods (e.g., Vanilla RAG) to state-of-the-art graph-based approaches (e.g., HippoRAG2, E2GraphRAG). The results are detailed in Table~\ref{tab:sub_results}.

\textbf{MemGraphRAG consistently achieves state-of-the-art performance across all evaluated datasets, highlighting its compatibility and robustness across different backbone models.} As shown in the table, MemGraphRAG achieves the highest average performance of 58.41\%, significantly outperforming the strongest baseline, HippoRAG2 (55.41\%), and surpassing standard graph-based methods like LightRAG (47.81\%) by a substantial margin. First, compared to non-structured methods, our approach exhibits a dominant advantage over Vanilla RAG (Top-5 average: 47.52\%), validating that our memory-driven graph structure effectively captures long-range dependencies that vector retrieval misses. Second, in the realm of graph-based RAG, MemGraphRAG excels particularly in multi-hop reasoning tasks. On the 2WikiMultiHopQA dataset, we achieve a Containment Accuracy of 69.40\% and an LLM Accuracy of 66.80\%, notably higher than HippoRAG2 (61.90\% and 54.40\%, respectively). This indicates that our method constructs a more connected and logically coherent graph, enabling the retriever to accurately locate multi-hop evidence chains. Furthermore, on domain-specific datasets like G-Medical, MemGraphRAG maintains its lead (67.13\%), proving its robustness in handling specialized knowledge. Collectively, these results confirm that MemGraphRAG provides a high-quality, globally consistent indexing structure that universally enhances the reasoning capabilities of LLMs.

\begin{table*}[t]
\centering
\caption{Comparison of different methods. The column \textbf{$\Delta$} shows the improvement of \textbf{MemGraphRAG} (58.41) over baselines. Darker green in $\Delta$ indicates a larger performance gap.}
\label{tab:sub_results}
\vspace{-2mm}
\resizebox{1\textwidth}{!}{
\begin{tabular}{lcccccccccc}
\toprule
\multirow{2}{*}{\textbf{Method}} 
    & \multicolumn{2}{c}{\textbf{HotpotQA}} 
    & \multicolumn{2}{c}{\textbf{2WikiMultiHopQA}}
    & \multicolumn{2}{c}{\textbf{MuSiQue}} 
    & \textbf{G-Medical}
    & \textbf{G-Novel}
    & \multicolumn{2}{c}{\textbf{Overall}} \\
\cmidrule(lr){2-3} \cmidrule(lr){4-5} \cmidrule(lr){6-7} \cmidrule(lr){8-8} \cmidrule(lr){9-9} \cmidrule(lr){10-11}
    & Contain-Acc. & LLM-Acc. & Contain-Acc & LLM-Acc & Contain-Acc & LLM-Acc & LLM-Acc & LLM-Acc & Avg. & $\Delta$ \\
\midrule
\multicolumn{11}{c}{\textbf{\textit{Non-structure Methods}}} \\
\midrule
llama-70B-instruct  & 38.20 & 40.50 & 33.80 & 29.30 & 14.40 & 16.50 & 39.31 & 29.73 & 30.22 & \cghigh{28.19} \\
Vanilla RAG (Top-1) & 48.40 & 51.70 & 40.20 & 34.40 & 22.20 & 23.10 & 52.73 & 45.13 & 39.73 & \cghigh{18.68} \\
Vanilla RAG (Top-3) & 54.50 & 55.10 & 46.60 & 39.20 & 29.50 & 31.70 & 55.38 & 48.90 & 45.11 & \cgmid{13.30} \\
Vanilla RAG (Top-5) & 56.00 & 58.10 & 53.20 & 46.70 & 28.20 & 32.30 & 59.81 & 45.83 & 47.52 & \cgmid{10.89} \\
\midrule
\multicolumn{11}{c}{\textbf{\textit{Graph-based RAG Methods}}} \\
\midrule
KGP                 & 63.20 & 62.10 & 34.70 & 32.60 & 24.80 & 30.20 & 56.94 & 47.04 & 43.95 & \cgmid{14.46} \\
G-retriever         & 44.70 & 43.70 & 50.50 & 30.80 & 18.90 & 19.80 & 50.77 & 43.09 & 37.78 & \cghigh{20.63} \\
RAPTOR              & 57.10 & 59.70 & 54.70 & 45.00 & 26.50 & 32.10 & 57.63 & 42.83 & 46.95 & \cgmid{11.46} \\
MS-GraphRAG         & 49.90 & 42.70 & 50.00 & 39.80 & 17.60 & 22.20 & 53.22 & 47.71 & 40.39 & \cghigh{18.02} \\
LazyGraphRAG        & 50.80 & 41.40 & 46.60 & 36.80 & 23.10 & 27.70 & 58.79 & 48.63 & 41.73 & \cghigh{16.68} \\
LightRAG            & 64.30 & 63.10 & 55.30 & 40.70 & 28.00 & 27.70 & 58.68 & 44.67 & 47.81 & \cgmid{10.60} \\
HippoRAG            & 59.50 & 64.00 & \underline{69.00} & \underline{63.80} & 31.80 & 28.40 & 57.30 & 47.44 & 52.66 & \cgmid{5.75} \\
HippoRAG2           & \underline{64.90} & \underline{67.30} & 61.90 & 54.40 & \textbf{33.90} & \underline{37.60} & \underline{67.11} & \textbf{56.16} & \underline{55.41} & \cglow{3.00} \\
$E^2$GraphRAG       & 61.50 & 65.70 & 58.60 & 40.50 & 24.20 & 27.40 & 62.47 & 53.82 & 49.27 & \cgmid{9.14} \\
\midrule
\rowcolor{ourrowbg}
\textbf{MemGraphRAG(ours)} & \textbf{65.60} & \textbf{69.40} & \textbf{69.40} & \textbf{66.80} & \underline{33.70} & \textbf{39.50} & \textbf{67.13} & \underline{55.76} & \textbf{58.41} & \textbf{--} \\
\bottomrule
\end{tabular}
}
\end{table*}
\subsection{Graph Analysis}
\begin{figure}[h] 
    \centering  
    \includegraphics[width=0.8\linewidth]{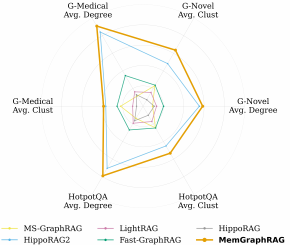} 
    \caption{Multi-dimensional assessment of graph quality.}
    \label{fig:graph}
\end{figure}

To more intuitively assess the quality of the index graphs produced by our memory-based construction approach, we analyze their topological properties and compare MemGraphRAG with existing baselines in terms of connectivity, redundancy, and semantic aggregation. Following previous study~\cite{xiang2025use}, we assessed the Average Degree and Average Clustering Coefficient of the index graphs constructed by various GraphRAG frameworks on the G-Medical and G-Novel datasets. The comparative results are presented in Table~\ref{tab:graph} and Figure~\ref{fig:graph}.

\textbf{MemGraphRAG demonstrates superior entity-level connectivity compared to existing GraphRAG methods.} MemGraphRAG achieves the highest Average Degree on both datasets, reaching 14.37 on the Medical dataset (surpassing HippoRAG2's 13.31) and 9.26 on the Novel dataset (surpassing HippoRAG2's 8.75). This improvement indicates that our memory consistency maintenance mechanism effectively links entities scattered across different document chunks. As a result, it bridges fragmented subgraphs and enables more robust long-range reasoning paths.

\textbf{MemGraphRAG demonstrates superior subgraph-level semantic clustering than existing GraphRAG methods.} MemGraphRAG also attains the highest Average Clustering Coefficients, with 0.865 on the G-Novel and 0.527 on the G-Medical. These results indicate that nodes in MemGraphRAG tend to share common neighbors, leading to denser local connectivity and clearer semantic clusters. This further shows that MemGraphRAG integrates dispersed knowledge into a more unified and highly structured index graph, instead of yielding sparse graphs composed of loosely related facts.

\begin{table}[t]
\caption{Quality evaluation of indexing graph construction in GraphRAG frameworks.}
\label{tab:graph}
\centering
\small
\resizebox{\linewidth}{!}{
\begin{tabular}{lcccccc}
\toprule
\multirow{2}{*}{\textbf{Method}} & \multicolumn{2}{c}{\textbf{G-Novel}} & \multicolumn{2}{c}{\textbf{G-Medical}} & \multicolumn{2}{c}{\textbf{HotpotQA}} \\
\cmidrule(lr){2-3} \cmidrule(lr){4-5} \cmidrule(lr){6-7} 
 & \textbf{Degree} & \textbf{Clust. Coeff} & \textbf{Degree} & \textbf{Clust. Coeff} & \textbf{Degree} & \textbf{Clust. Coeff} \\
\midrule
\textbf{MS-GraphRAG}~\cite{edge2024local}     & 1.48 & 0.315 & 1.82  & 0.300 & 1.56 & 0.334 \\
\textbf{HippoRAG2}~\cite{gutiérrez2025hipporag2}   & \underline{8.75} & \underline{0.657} & \underline{13.31} & \underline{0.497} & \underline{7.96} & \underline{0.613} \\
\textbf{LightRAG}~\cite{guo2024lightrag}      & 2.10  & 0.212 & 2.58  & 0.139 & 2.18 & 0.236 \\
\textbf{Fast-GraphRAG}~\cite{fastgraphrag2024}      & 3.19 & 0.324 & 5.50   & 0.347 & 3.04 & 0.336 \\
\textbf{HippoRAG}~\cite{hipporag}               & 1.73 & 0.100   & 2.06  & 0.087 & 1.86 & 0.140 \\
\midrule
\textbf{MemGraphRAG(ours)}                   & \textbf{9.26} & \textbf{0.865} & \textbf{14.37} & \textbf{0.527} & \textbf{8.92} & \textbf{0.725} \\
\bottomrule
\end{tabular}}
\end{table}

\subsection{Case Study}
We conduct a qualitative analysis in Table~\ref{tab:case_conflict} and Table~\ref{tab:case_denoising} to illustrate how MemGraphRAG overcomes the limitations of isolated extraction by ensuring logical consistency and thematic purity through its global memory mechanism.

1) \textbf{Case Study on Conflict Resolution.} Table~\ref{tab:case_conflict} illustrates a representative scenario of Mutually Exclusive Conflict, where disparate documents claim conflicting birth years for the same entity ("1645" vs. "1643"). Traditional pipelines simply aggregate these contradictions, leading to ambiguous reasoning paths. MemGraphRAG addresses this through Global Adjudication. Upon detecting the conflict, the Resolution Agent ($A_{res}$) retrieves the original provenance from the Passage Layer ($M_{pas}$) and validates the correct fact ("1643") before indexing. This mechanism effectively eliminates logical incoherence, enabling the retriever to provide an accurate context for the LLM.

2) \textbf{Case Study on Thematic Denoising.} In domain-specific tasks (e.g., medical protocols), LLMs often extract irrelevant noise alongside core facts. As shown in Table~\ref{tab:case_conflict}, the baseline graph is polluted by irrelevant triples (e.g., Patient prefers Tea), which distracts the retrieval process. MemGraphRAG mitigates this via Unified Schema Filtering. By treating extracted schemas as candidate and only stabilizing those that exceed a frequency threshold ($\tau$), our system successfully filters out irrelevant noise while retaining stable clinical patterns (e.g., $Drug \xrightarrow{treats} Disease$). This results in a cleaner Fact Graph ($G_{fac}$) that strictly follows the domain ontology, significantly improving retrieval precision.

\begin{table*}[t]
\centering
\caption{Case Study: Resolving Logic Conflicts via Global Adjudication. Comparing how MemGraphRAG handles contradictory birth years across documents versus a Traditional GraphRAG baseline.}
\label{tab:case_conflict}
\small
\begin{tabularx}{\textwidth}{l|X|X}
\toprule
\textbf{Pipeline Phase} & \textbf{Traditional GraphRAG (Baseline)} & \textbf{MemGraphRAG (Ours)} \\
\midrule
\textbf{1. Input Corpus} & Doc A: \textit{``Newton was born in 1645.''} \newline Doc B: \textit{``Isaac Newton, born 1643...''} & \textbf{Same Corpus}: Contains mutually exclusive facts due to source errors or extraction noise. \\
\midrule
\textbf{2. Graph Construction} & \textbf{Isolated Extraction}: \newline $T_1: (Newton, born\_in, 1645)$ \newline $T_2: (Newton, born\_in, 1643)$ \newline $\rightarrow$ \textit{Both edges added to Graph $G$.} & \textbf{Global Adjudication}: \newline $A_{det}$ detects Conflict: $T_1 \perp T_2$ \newline $\rightarrow A_{res}$ checks Evidence ($M_{pas}$) \newline $\rightarrow$ \textbf{Update}: Keep $T_2$, Discard $T_1$. \\
\midrule
\textbf{3. Retrieval Query} & \multicolumn{2}{c}{\textbf{Q: ``When was Isaac Newton born?''}} \\
\midrule
\textbf{4. Retrieval Process} & \textbf{Noisy Activation}: \newline Query triggers both nodes: $\{1645, 1643\}$ \newline $\rightarrow$ Retriever fetches conflicting context. & \textbf{Consistent Path}: \newline Query triggers verified node: $\{1643\}$ \newline $\rightarrow$ Trace back to $M_{pas}$ evidence. \\
\midrule
\textbf{5. Final Answer} & \textit{``Newton was born in 1645 or 1643...''} \newline (\textbf{Ambiguous / Hallucinated}) & \textit{``Isaac Newton was born in 1643.''} \newline (\textbf{Precise \& Verified}) \\
\bottomrule
\end{tabularx}
\end{table*}

\begin{table*}[t]
\centering
\caption{Case Study: Thematic Denoising in Medical Protocols. Demonstrating how MemGraphRAG filters irrelevant extraction noise using Unified Schema Filtering.}
\label{tab:case_denoising}
\small
\begin{tabularx}{\textwidth}{l|X|X}
\toprule
\textbf{Pipeline Phase} & \textbf{Traditional GraphRAG (Baseline)} & \textbf{MemGraphRAG (Ours)} \\
\midrule
\textbf{1. Input Corpus} & Chunk 1: \textit{``Osimertinib treats EGFR-mutant NSCLC.''} \newline Chunk 2: \textit{``Patient prefers tea over coffee.''} & \textbf{Same Corpus}: Mixture of clinical facts and irrelevant patient anecdotes. \\
\midrule
\textbf{2. Graph Construction} & \textbf{Full Extraction}: \newline $T_1: (Osimertinib, treat, NSCLC)$ \newline $T_2: (Patient, prefer, Tea)$ \newline $\rightarrow$ \textit{Noise $T_2$ pollutes the graph.} & \textbf{Schema Filtering}: \newline Schema $S_1 (Drug, treat, Dis)$ freq $\ge \tau \rightarrow$ \textbf{Stable} \newline Schema $S_2 (Pat, pref, Bev)$ freq $< \tau \rightarrow$ \textbf{Pending} \newline $\rightarrow$ \textbf{Result}: Only $T_1$ activated in $G_{fac}$. \\
\midrule
\textbf{3. Retrieval Query} & \multicolumn{2}{c}{\textbf{Q: ``What is the standard treatment for NSCLC?''}} \\
\midrule
\textbf{4. Retrieval Process} & \textbf{Drifting Path}: \newline Node \textit{NSCLC} $\rightarrow$ \textit{Patient} $\rightarrow$ \textit{Tea} \newline $\rightarrow$ Retrieves irrelevant dietary info. & \textbf{Focused Path}: \newline Node \textit{NSCLC} $\rightarrow$ \textit{Osimertinib} \newline $\rightarrow$ Strictly follows clinical ontology. \\
\midrule
\textbf{5. Final Answer} & \textit{``Osimertinib is used. Patients may prefer tea.''} \newline (\textbf{Unprofessional / Distracted}) & \textit{``Osimertinib is the recommended treatment.''} \newline (\textbf{Professional \& Concise}) \\
\bottomrule
\end{tabularx}
\end{table*}

\section{Related Work}

\subsection{Retrieval-Augmented Generation} While Large Language Models (LLMs) have demonstrated impressive capabilities, they remain prone to hallucination~\citep{fang2024alphaedit,jiang2025anyedit,fang2025safemlrm,zheng2025usb,hong2024next,yuan2025knapsack,zhong2024iterative,dong2026use,gao2025probing,lin2026zerounlearnfewshotknowledgeunlearning,lin2026bidirectional}. Retrieval-Augmented Generation (RAG) mitigates this by grounding generation in external evidence~\citep{guu2020realmretrievalaugmentedlanguagemodel,borgeaud2022improving,izacard2023atlas,qian2025memoragboostinglongcontext,zhou2025improving,zhou2025essence,zhang2025erarag}. However, effectively organizing fragmented knowledge from distributed documents to support complex reasoning remains a persistent challenge.

To address this, recent research has evolved from simple retrieval to Reasoning-enhanced RAG~\cite{asai2023self,li2025searcho1agenticsearchenhancedlarge,jin2025searchr1trainingllmsreason,chen2025researchlearningreasonsearch,xiang2026systematic}. Departing from static index construction, this paradigm focuses on interleaving the retrieval process with the logical flow of the LLM. Several approaches optimize the retrieval process through Chain-of-Thought prompting, recursive inner monologues, or logical decomposition, such as IRCoT~\citep{trivedi2023interleaving}, IM-RAG~\citep{yang2024rag}, and LAG~\citep{xiao2025lag}. LogicRAG~\citep{chen2025logicrag} advances this direction by eliminating pre-built graphs entirely, instead constructing a reasoning Directed Acyclic Graph (DAG) dynamically at inference time to enable adaptive retrieval planning. While effective, these methods typically operate within the constraints of fixed resources or rely on the LLM's inherent reasoning capabilities rather than structured knowledge representation.

\subsection{Graph Retrieval-Augmented Generation} To overcome the limitations of unstructured text chunks, GraphRAG focuses on explicit graph structure construction to capture global dependencies and structural patterns. Current approaches can be categorized into two primary construction paradigms:

\textbf{Relation-extraction-based Construction.} This line of work~\citep{hipporag,guo2024lightrag,xiao2025reliablereasoningpathdistilling,yang2025graphsearchagenticdeepsearching,gusarov2025multiagentgraphragtexttocypherframework,zhao20252graphrag,zhou2025depth,luo2026utility,tsang2025autographr1endtoendreinforcementlearning,chen2025logicrag} structures text corpora into Knowledge Graphs (KGs) by extracting triples to form atomic knowledge units. These units are subsequently unified via entity alignment~\citep{neusymea, LLM4EA}, enabling the application of sophisticated graph reasoning algorithms~\citep{pLogicNet, sun2024thinkongraph, luo2024reasoning}. Some methods augment reasoning by integrating these static KGs as navigational aids, such as Think-on-Graph~\citep{sun2024thinkongraph} and RRP~\citep{xiao2025reliablereasoningpathdistilling}. However, independent OpenIE extraction often leads to inconsistency. Although schema-guided approaches~\citep{liang2024kag,sharma2024og} attempt to standardize this, they entail high manual costs. Addressing these inefficiencies, LinearRAG~\citep{zhuang2025linearrag} proposes a relation-free ``Tri-Graph'' based on lightweight entity extraction, achieving linear scalability without the noise associated with traditional triple extraction.

\textbf{Clustering-based Hierarchy Construction.} Complementary to triple-based methods, this category focuses on capturing global information by identifying dense structural patterns. Methods typically employ community detection algorithms, such as Louvain or Leiden, to recursively aggregate entities into clusters~\citep{edge2024local,sarthi2024raptor, hipporag}. These clusters serve as hierarchical summaries, abstracting raw passages into topic-level communities to provide a macro-level perspective. Despite its utility in summarizing high-level themes, this unsupervised approach faces limitations regarding precision, as inaccuracies in low-level entity relationships can propagate upward, and the iterative clustering of large-scale graphs poses significant bottlenecks for real-time deployment.

\section{Details of Preliminary Study}\label{app:pre}
Independent extraction across different chunks may introduce conflicting information into the merged graph, resulting in semantic contradictions. In our preliminary study, we identify three major types of such conflicts, as summarized in Table~\ref{tab:knowledge_conflicts}. Specifically:
\begin{itemize}
    \item \textbf{Mutually Exclusive Conflict:} Facts that cannot coexist in reality. For example, Chunk A yields \textit{(Newton, Birth year, 1643)} while Chunk B yields \textit{(Newton, Birth year, 1645)}.
    \item \textbf{Temporal Conflict:} Contradictions arising from time-variant facts. A corpus spanning different years may generate both \textit{(Biden, President, USA)} and \textit{(Trump, President, USA)} without temporal qualifiers, confusing the retriever.
    \item \textbf{Granularity Conflict:} Facts describing the same reality at incompatible abstraction levels. For instance, connecting an entity to both specific and general concepts, such as \textit{(Xiao Ming, born\_in, Shanghai)} and \textit{(Xiao Ming, born\_in, China)}, or \textit{(AI, subclass, NLP)} vs. \textit{(AI, subclass, Unsupervised Learning)}. These inconsistencies create redundant paths that dilute the reasoning focus.
\end{itemize}

\begin{table*}[h]
\centering
\caption{Taxonomy of Knowledge Conflicts in Graph Retrieval-Augmented Systems.}
\label{tab:knowledge_conflicts}
\renewcommand{\arraystretch}{1.5} 
\begin{tabular}{p{3cm}|p{3.5cm}|p{4.5cm}|p{4cm}}
\hline
\textbf{Conflict Type} & \textbf{Definition} & \textbf{Mechanism \& Impact} & \textbf{Illustrative Examples} \\ \hline
\textbf{Mutually Exclusive Conflict} & 
Logically incompatible facts that cannot simultaneously hold true within a single domain of discourse. & 
\textbf{Mechanism:} Distinct sources attribute divergent values to a functional property (single-value attribute). \newline
\textbf{Impact:} Introduces binary logical contradictions that halt deterministic reasoning. & 
\textit{Attribute Value Clash:} \newline
Source A: \textit{(Newton, born\_in, 1643)} \newline
Source B: \textit{(Newton, born\_in, 1645)} \\ \hline

\textbf{Temporal Conflict} & 
Inconsistencies arising from state changes in time-variant facts when temporal metadata is absent. & 
\textbf{Mechanism:} Facts valid in disjoint time intervals ($T_1 \neq T_2$) are flattened into a static knowledge base. \newline
\textbf{Impact:} Confuses the retriever by presenting outdated or competing truths as currently valid. & 
\textit{Role Evolution:} \newline
$T_{2020}$: \textit{(Trump, President, USA)} \newline
$T_{2021}$: \textit{(Biden, President, USA)} \newline
\textit{(Both retrieved without timestamps)} \\ \hline

\textbf{Granularity Conflict} & 
Discrepancies in the level of abstraction or specificity regarding the same entity or concept. & 
\textbf{Mechanism:} Simultaneous mapping of an entity to hierarchically distinct nodes (e.g., specific vs. general) within an ontology. \newline
\textbf{Impact:} Creates redundant inference paths and dilutes reasoning precision. & 
\textit{Geospatial:} \newline
\textit{(Xiao Ming, born\_in, Shanghai)} vs. \textit{(Xiao Ming, born\_in, China)} \newline
\textit{Taxonomical:} \newline
\textit{(AI, subclass, NLP)} vs. \textit{(AI, subclass, Machine Learning)} \\ \hline
\end{tabular}
\end{table*}

\section{Details of the Proposed Method}
\subsection{Key Definitions}
To establish a rigorous foundation for the subsequent methodology, we first provide formal definitions for the core components of our hierarchical knowledge representation:

(i) \textbf{Type ($t$) and Entity ($e$)}: We distinguish between abstract concepts and concrete instances. A type $t \in \mathcal{T}$ denotes a high-level taxonomic category (e.g., \textit{Person}) that serves as a semantic anchor. An entity $e \in \mathcal{E}$ refers to a specific instance grounded in the text (e.g., \textit{Einstein}), where each entity is associated with a type through a mapping function $\phi(e)=t$.

(ii) \textbf{Schema ($s$) and Fact ($f$)}: We define knowledge triples at two levels of abstraction. A schema $s=(t_h, r, t_t)$ specifies a structural constraint, where $t_h, t_t \in \mathcal{T}$ represent the head and tail types, and $r$ denotes a semantic relation (e.g., (\textit{Person}, \textit{born\_in}, \textit{Country})). A fact $f=(e_h, r, e_t)$ is a concrete instantiation of a schema, where $e_h, e_t \in \mathcal{E}$ (e.g., (\textit{Einstein}, \textit{born\_in}, \textit{Germany})).

(iii) \textbf{Ontology ($\mathcal{O}$)}: The ontology is defined as the structured collection of all valid schemas, denoted as $\mathcal{O}=\{s_1,\dots,s_n\}$. It governs the structural rules of the knowledge graph by enforcing semantic constraints, ensuring that all extracted facts conform to predefined schema specifications.

(iv) \textbf{Passage ($p$)}: A passage $p \in \mathcal{P}$ represents a granular segment of raw text from the corpus, serving as the evidence grounding unit. Specifically, each extracted fact $f$ is explicitly linked to its supporting textual evidence through a mapping function $\psi(f) \to p_i$.

\subsection{MemGraphRAG architecture}\label{sec3.1}
To overcome fragmented extraction and support the coherent evolution of knowledge graphs, we propose the MemGraphRAG architecture. Our core premise is that high-quality graph construction requires not only structured storage, but also active management of knowledge. The system is built upon two complementary components: (i) a \textit{Hierarchical Memory Architecture} that organizes schemas, facts, and passages across different abstraction levels, and (ii) a \textit{Multi-Agent System} that serves as the dynamic execution engine, leveraging memory to drive the iterative ``extract--verify--modify'' process. In the following sections, we describe how these components work together to ensure global consistency.

\label{appendix:indexing_mech}
\textbf{Global Memory}, which organizes knowledge in a three-tier structure that aligns abstract schemas, concrete facts, and supporting evidence. The top-level \textit{Ontology Layer} ($\mathcal{M}_{ont}$) maintains schema patterns with their statistical frequencies, providing semantic structure and global theme for graph construction. The middle \textit{Fact Layer} ($\mathcal{M}_{fac}$) stores instantiated triples derived from these schemas. The lowest \textit{Passage Layer} ($\mathcal{M}_{pas}$) preserves the original source passages, ensuring that extracted facts remain grounded in their linguistic context. 

To strengthen associations across layers, we introduce a \textbf{dense indexing mechanism} that enforces structural consistency through bidirectional interactions. Specifically, \textit{Schema--Instance Alignment} is established not merely as a one-way classification, but as a mutual binding between abstraction and instantiation. On the bottom-up direction, we define a mapping
\begin{equation}
\Phi : \mathcal{M}_{fac} \rightarrow \mathcal{M}_{ont},
\end{equation}
which enforces strict typing by assigning each triple $t \in \mathcal{M}_{fac}$ to a schema constraint $s \in \mathcal{M}_{ont}$. On the top-down direction, each schema $s$ induces its instantiation set
\begin{equation}
\mathcal{T}(s) = \{\, t \in \mathcal{M}_{fac} \mid \Phi(t) = s \,\},
\qquad |\mathcal{T}(s)| \ge 0,
\end{equation}
capturing the duality that schemas constrain facts while facts substantiate schemas.

Simultaneously, \textit{Fact--Evidence Grounding} is modeled via a bidirectional relation
\begin{equation}
\Psi \subseteq \mathcal{M}_{fac} \times \mathcal{M}_{pas},
\end{equation}
which links each fact to its supporting passages (provenance) while allowing passages to index the facts they yield (extraction). For any triple $t$, we define its evidence set as
\begin{equation}
\mathcal{E}(t) = \{\, p \in \mathcal{M}_{pas} \mid (t, p) \in \Psi \,\},
\qquad |\mathcal{E}(t)| \ge 1.
\end{equation}
Together, these bidirectional mappings ensure that the graph is both logically governed by the ontology and rigorously grounded in textual evidence.

\label{appendix:hierar_graph}
\textbf{Hierarchical Indexing Graph}, which provides a unified representation spanning abstract schemas, concrete facts, and textual evidence. Concretely, we organize $\mathcal{G}$ into three interconnected graph views that enable hierarchical navigation from high-level semantic concepts to fine-grained supporting passages. (i) \textit{Semantic Ontology Graph ($\mathcal{G}_{ont}$)}: Derived from the ontology layer $\mathcal{M}_{ont}$, $\mathcal{G}_{ont}$ forms a high-level network of domain types and schema relations. It serves as the logical backbone of the overall graph by encoding valid relational patterns and domain constraints. (ii) \textit{Fact Graph ($\mathcal{G}_{fac}$)}: Constructed from the fact layer $\mathcal{M}_{fac}$, $\mathcal{G}_{fac}$ represents an entity-relation graph over instantiated triples, which acts as the primary substrate for multi-hop reasoning. (iii) \textit{Source Evidence Graph ($\mathcal{G}_{pas}$)}: Induced from the passage layer $\mathcal{M}_{pas}$, $\mathcal{G}_{pas}$ grounds entities and relations in $\mathcal{G}_{fac}$ back to their originating text passages, providing fine-grained evidence support for faithful answer generation. Together, this multi-view architecture enables structured reasoning that progressively traverses from $\mathcal{G}_{ont}$ to $\mathcal{G}_{fac}$, and finally to $\mathcal{G}_{pas}$ for evidence retrieval.

\textbf{Multi-Agent System}, which introduces the dynamic execution units that drive the system's evolution, is formulated as a collaborative ecosystem of specialized agents interacting with $\mathcal{M}$ through distinct cognitive roles. Specifically, the Multi-Agent System is defined as $\mathcal{A} = {A_{ext}, A_{det}, A_{res}}$, where each agent focuses on a separate function. Our design philosophy emphasizes the decoupling of generation, diagnosis, and correction to ensure high-fidelity graph construction: (i) the \textbf{Extraction Agent} ($A_{ext}$), which initializes the graph by processing input documents and populating all three layers of $\mathcal{M}$ (Schema, Fact, and Passage) in parallel, ensuring that each extracted fact is grounded in supporting evidence; (ii) the \textbf{Conflict Detection Agent} ($A_{det}$), which is triggered by updates in the Fact Layer ($\mathcal{M}{fac}$) and performs purely diagnostic checks to identify structural anomalies, redundancy, and logical inconsistencies; and (iii) the \textbf{Conflict Resolution Agent} ($A{res}$), which resolves conflicts flagged by $A_{det}$ by leveraging the global context stored in $\mathcal{M}$, including historical evidence in $\mathcal{M}{pas}$ and schema constraints in $\mathcal{M}{ont}$, thereby maintaining the global consistency of $\mathcal{G}$.

\begin{algorithm*}[t]
\caption{Memory-based Indexing Graph Construction}
\label{alg:graph_construction}
\begin{algorithmic}[1]

\Require Document stream $\mathcal{D}$; Global memory $\mathcal{M}=\{M_{ont},M_{fac},M_{pas}\}$;
schema threshold $\tau$; conflict threshold $\delta$; bridging threshold $\delta_b$
\Ensure Global hierarchical graph $\mathcal{G}$

\vspace{0.2em}
\State \textbf{Stage I: Composite Extraction into Memory (Sandbox)}
\For{each chunk $c_i$ from $\mathcal{D}$}
    \State $\{O_{cand}, T_{cand}, P_{src}\} \gets A_{ext}(c_i)$
    \Comment{Extract candidate schemas, triples, and provenance}
    \State Store $O_{cand},T_{cand},P_{src}$ into $(M_{ont},M_{fac},M_{pas})$
    \Comment{Probationary storage: extraction as hypotheses}
\EndFor

\vspace{0.2em}
\State \textbf{Stage II: Unified Schema Filtering and Triple Activation}
\For{each schema $o \in M_{ont}$}
    \If{$\mathrm{Freq}(o) \ge \tau$}
        \State $\mathrm{State}(o)\gets \texttt{Stable}$
        \Comment{Promote only consensus schemas}
    \EndIf
\EndFor
\For{each triple $t \in M_{fac}$}
    \If{$\mathrm{State}(\textsc{Schema}(t))=\texttt{Stable}$}
        \State $\mathrm{State}(t)\gets \texttt{Active}$
        \Comment{Activate only triples governed by stable schema}
    \EndIf
\EndFor

\vspace{0.2em}
\State \textbf{Stage III: Conflict Detection and Evidence-based Adjudication}
\For{each newly active triple $t_{new}$}
    \State $\mathcal{F}_{conf} \gets \{t' \in M_{fac} \mid \text{Sim}(t_{new},t')>\delta \ \lor\ \text{Match}(t_{new},t')\}$
    \Comment{Global scan for logical/temporal/granularity conflicts}
    \If{$\mathcal{F}_{conf}\neq \emptyset$}
        \State $C_{ctx} \gets \Psi(t_{new}) \cup \bigcup_{t'\in \mathcal{F}_{conf}}\Psi(t')$
        \Comment{Retrieve provenance passages as evidence}
        \State $A_{res}$ updates $t_{new}$ and $\mathcal{F}_{conf}$ based on $C_{ctx}$
        \Comment{Discard / refine / temporally augment conflicting facts}
    \EndIf
\EndFor

\vspace{0.2em}
\State \textbf{Stage IV: Multi-view Projection and Memory-guided Bridging}
\State Construct $\mathcal{G}_{ont}$ from stable schemas in $M_{ont}$
\State Construct $\mathcal{G}_{fac}$ from active triples in $M_{fac}$
\State Construct $\mathcal{G}_{pas}$ from provenance passages in $M_{pas}$
\Comment{Project memory layers into graph views}

\State Add type-based edges linking entities with shared schema types
\Comment{Type-based bridging for disjoint subgraphs}
\State Add similarity-based edges if $\text{Sim}(e_i,e_j)>\delta_b$
\Comment{Embedding-based bridging for long-range connectivity}

\State Merge all views into global hierarchical graph $\mathcal{G}$
\State \Return $\mathcal{G}$

\end{algorithmic}
\end{algorithm*}

\subsection{Memory-based Indexing Graph Construction}
Traditional graph construction often processes document chunks in isolation, leading to redundant entities and fragmented subgraphs. To address this, we reframe graph construction not as a one-off extraction task, but as a \textbf{dynamic co-evolution process} between the Global Memory $\mathcal{M}$ and the Knowledge Graph $\mathcal{G}$. 
Driven by the memory system, we implement two strategic paradigms to ensure structural integrity:
(i) \textbf{Structure Optimization via Progressive Construction}: Instead of trusting LLM outputs immediately, we treat extractions as hypotheses. The memory acts as a ``probationary sandbox,'' allowing the graph to evolve via an iterative ``extract–verify–modify'' cycle that filters noise before it pollutes the graph structure.
(ii) \textbf{Conflict Resolution via Global Perspective}: By maintaining a persistent global state, our shared memory enables the system to detect and resolve semantic contradictions (e.g., logical, temporal, or granular conflicts) that span across disparate documents, ensuring a unified and consistent knowledge representation.

\subsubsection{Thematic Denoising via Unified Schema Filtering}\label{appendix:dinoising}
To mitigate the stochastic hallucinations inherent in LLMs and ensure statistical consensus, we implement a \textbf{``Probationary Extraction Protocol.''} This protocol enforces a strict separation between raw extractions and validated knowledge.

\textbf{First, Composite Extraction into Memory.} 
The process initiates by partitioning the document stream into uniform chunks $c_i \in \mathcal{C}$. For each chunk, the Extraction Agent ($A_{ext}$) generates a \textit{Composite Extraction Record} that simultaneously populates all three memory layers:
\begin{equation}
A_{ext}(c_i) \rightarrow \{O_{\text{cand}}, T_{\text{cand}}, P_{\text{src}}\}
\end{equation}
where $O_{\text{cand}}$ and $T_{\text{cand}}$ represent candidate schemas and triples, and $P_{\text{src}}$ anchors them to the source text.

\textbf{Second, The Ontology Filter Mechanism.} 
Crucially, newly extracted schemas are initially assigned a logical ``Candidate State'' (Pending). While physically stored in memory for tracking, they remain \textit{invisible} to the global graph structure $\mathcal{G}$. This isolation prevents low-frequency noise from polluting the index.

\textbf{Finally, Confidence-Driven State Promotion.} 
We formalize the evolution of knowledge using a frequency-based confidence function. A schema transitions from ``Pending'' to ``Stable'' only when its extraction frequency across the corpus exceeds a statistical threshold $\tau$:
\begin{equation}
\mathrm{State}(o)=
\begin{cases}
\texttt{Stable}, & \text{if } \mathrm{Freq}(o) \ge \tau, \\[6pt]
\texttt{Pending}, & \text{otherwise}.
\end{cases}
\end{equation}
This transition triggers a \textit{cascading activation}: only triples governed by a stable schema are flagged as ``Active.'' Only these active triples are permitted to enter the subsequent conflict detection phase, ensuring the graph is constructed solely from consensus-verified knowledge.

\subsubsection{Consistency Maintenance via Global Adjudication}
Dynamic graph updates inevitably introduce contradictions. To ensure trustworthiness, we implement a collaborative mechanism where agents utilize Global Memory as the ``ground truth'' for adjudication.

\textbf{Step 1: Asynchronous Conflict Triggering.} 
The Conflict Detection Agent ($A_{det}$) is triggered strictly when a triple $t_{\text{new}}$ transitions to an ``Active'' state. $A_{det}$ performs a hybrid scan over the existing Fact Memory ($\mathcal{M}_{\text{fac}}$), utilizing both vector similarity and symbolic matching to identify potential conflict candidates $T_{\text{conf}}$:
\begin{equation}
T_{\mathrm{conf}} = \left\{ t' \in \mathcal{M}_{\mathrm{fac}} \;\middle|\; \mathrm{Sim}(t_{\mathrm{new}}, t') > \delta \;\lor\; \mathrm{Match}(t_{\mathrm{new}}, t') \right\}.
\end{equation}
If $T_{\mathrm{conf}} \neq \emptyset$, the resolution protocol is initiated.

\textbf{Step 2: Evidence Retrieval and Adjudication.} 
Unlike black-box resolution, our approach is evidence-driven. The Conflict Resolution Agent ($A_{res}$) leverages the memory mapping $\Psi$ to retrieve the original provenance for both the new assertion and the conflicting facts. It constructs a context window $C_{\text{ctx}}$ containing the raw source passages:
\begin{equation}
C_{\mathrm{ctx}} = \Psi(t_{\mathrm{new}}) \cup \bigcup_{t' \in T_{\mathrm{conf}}} \Psi(t').
\end{equation}
Based on $C_{\text{ctx}}$, $A_{res}$ reasons to determine factual validity, effectively acting as a judge reviewing case files.

\textbf{Step 3: Taxonomy-Based Resolution Strategies.} 
Based on the evidence, $A_{res}$ executes targeted updates to resolve specific conflict types:
\begin{itemize}
    \item \textit{Mutually Exclusive Conflict (Logical):} For contradictory facts (e.g., conflicting birthplaces), the agent compares evidence reliability to discard the erroneous fact.
    \item \textit{Temporal Conflict:} For facts valid in different periods (e.g., distinct presidential terms), the agent resolves ambiguity by appending temporal attributes (e.g., adding ``46th'' vs. ``47th'').
    \item \textit{Granularity Conflict (Structural):} For facts describing the same reality at different abstraction levels (e.g., ``Shanghai'' vs. ``China''), the agent refines predicates to allow logical coexistence (e.g., \textit{born\_city} vs. \textit{born\_country}).
\end{itemize}

\subsubsection{Structural Unification via Memory-Guided Bridging}
The final phase transforms the validated contents of the memory system into a navigable Global Hierarchical Graph $\mathcal{G}$. We adopt a \textbf{multi-view projection strategy} that maps the three memory layers into corresponding graph views: $\mathcal{G}_{ont}$ (Schema View), $\mathcal{G}_{fac}$ (Fact View), and $\mathcal{G}_{pas}$ (Source View).

To address the common issue of disjoint subgraphs in extracted knowledge, we augment the primary reasoning substrate, $\mathcal{G}_{fac}$, with two \textbf{memory-enabled connectivity mechanisms}:
\begin{enumerate}
    \item \textbf{Type-Based Bridging:} Leveraging $\mathcal{M}_{ont}$, disjoint entities are explicitly connected if they map to the same high-level schema type (e.g., connecting all \textit{Researchers} regardless of their document origin).
    \item \textbf{Similarity-Based Bridging:} Leveraging embedding storage in $\mathcal{M}$, we introduce implicit edges between entity pairs whose vector similarity exceeds a threshold $\delta$.
\end{enumerate}
These mechanisms leverage the global nature of memory to connect long-distance entities, significantly enhancing the graph's ability to support multi-hop reasoning across documents where explicit textual links are missing.

\begin{algorithm*}[t]
\caption{Memory-guided Online Retrieval}
\label{alg:retrieval}
\begin{algorithmic}[1]

\Require Query embedding $\mathbf{q}$; Graph $\mathcal{G}$ with transition matrix $\mathbf{M}$;
Memory $\mathcal{M}=\{M_{ont},M_{fac},M_{pas}\}$;
top-$K$; threshold $\tau$; damping $\lambda$; balance $\alpha$
\Ensure Evidence set $\mathcal{C}$ for downstream LLM generation

\vspace{0.2em}
\State \textbf{Stage I: Multi-layer Retrieval and Filtering}
\State $\mathcal{S}_{raw}\gets \textsc{TopK}(M_{ont},\mathbf{q},K)$;\ \ $\mathcal{F}_{raw}\gets \textsc{TopK}(M_{fac},\mathbf{q},K)$;\ \ $\mathcal{P}_{raw}\gets \textsc{TopK}(M_{pas},\mathbf{q},K)$
\Comment{Align query with ontology, facts, and passages}
\State $\mathcal{S}_{ret}\gets \{s\in\mathcal{S}_{raw}\mid \text{Sim}(\mathbf{q},\mathbf{s})>\tau\}$;\ \ $\mathcal{F}_{ret}\gets \{f\in\mathcal{F}_{raw}\mid \text{Sim}(\mathbf{q},\mathbf{f})>\tau\}$

\If{$\mathcal{S}_{ret}\cup \mathcal{F}_{ret}=\emptyset$}
    \State \Return $\mathcal{P}_{raw}$ \Comment{Fallback to standard RAG}
\EndIf

\vspace{0.2em}
\State \textbf{Stage II: Structure-aware Node Initialization}
\State Define reset weights $P_{init}(v)$ on nodes $v\in\mathcal{G}$:

\State \textit{Entity nodes:}\quad
$P_{init}(e)=\frac{1}{|\mathcal{F}_e|}\sum_{f\in\mathcal{F}_e}\text{Sim}(\mathbf{q},\mathbf{f}),\ 
\mathcal{F}_e=\{f\in\mathcal{F}_{ret}\mid e\in f\}$ \Comment{Ground by query-relevant facts}

\State \textit{Type nodes:}\quad
$P_{init}(t)=\left(\frac{1}{|\mathcal{S}_t|}\sum_{s\in\mathcal{S}_t}\text{Sim}(\mathbf{q},\mathbf{s})\right)\cdot\frac{1}{\log(\deg(t)+1)},\
\mathcal{S}_t=\{s\in\mathcal{S}_{ret}\mid t\in s\}$ \Comment{Schema relevance + hub suppression}

\State \textit{Passage nodes:}\quad
$P_{init}(p)=\text{Sim}(\mathbf{q},\mathbf{d}_p)\cdot \alpha \cdot
\sigma\!\left(\frac{\sum_{e\in\mathcal{E}_p}\text{IDF}(e)}{\log(|\mathcal{E}_p|+1)}\right)$
\Comment{Semantic alignment + information density}

\State Normalize $P_{init}$ into $\mathbf{p}^{(0)}$ with $\sum_v \mathbf{p}^{(0)}(v)=1$

\vspace{0.2em}
\State \textbf{Stage III: PPR Propagation and Evidence Selection}
\Repeat
    \State $\mathbf{p}^{(k+1)} \gets (1-\lambda)\mathbf{M}\mathbf{p}^{(k)} + \lambda \mathbf{p}^{(0)}$
    \Comment{Personalized PageRank with restart}
\Until{convergence}

\State Select top-ranked passages $\mathcal{P}^*$ and entities $\mathcal{E}^*$ by $\mathbf{p}^{(\infty)}$
\State $\mathcal{C}\gets \mathcal{P}^*\cup \mathcal{E}^*$
\State \Return $\mathcal{C}$

\end{algorithmic}
\end{algorithm*}
\definecolor{promptGreen}{RGB}{120, 195, 120} 
\definecolor{promptBlue}{RGB}{110, 120, 190}  

\newtcolorbox{mypromptbox}[2][]{
    enhanced,                 
    drop shadow,              
    colframe=#2,              
    colback=white,            
    coltitle=white,           
    title=\textbf{\Large #1}, 
    fontupper=\ttfamily,      
    sharp corners,            
    boxrule=1.5mm,            
    left=3mm, right=3mm, top=2mm, bottom=2mm, 
    arc=0mm,
    outer arc=0mm
}

\section{Prompt Set}\label{app:case}
To provide a more intuitive illustration of our graph construction procedure and ensure reproducibility, we present the \textit{Conflict Detection} and \textit{Conflict Resolution} components used in MemGraphRAG indexing, as shown in Figure~\ref{appendix:prompt1} and ~\ref{appendix:prompt2}.

\begin{figure*}[t!] 
\begin{mypromptbox}[Conflict Detection]{promptBlue}
\textbf{Task Definitions:} You are an expert fact checker. Given a target triple and a list of related triples. Your task: Detect whether target triple conflicts with any triple in the list of related triples, and classify conflicts into three types:
\begin{itemize}
    \item mutual conflict (mutual exclusivity / one-to-one relations)
    \item Temporal conflict (time-dependent facts; conflicts arise when time scopes overlap or are missing)
    \item Granularity conflict (different levels of specificity; may be compatible via containment)
\end{itemize}

Definitions and rules:

\textbf{1. mutual conflict (type = ``mutual")}
A mutual conflict happens when:
\begin{itemize}
    \item Same subject and predicate, but different objects, AND the predicate is one-to-one / mutually exclusive. Example: (X, birthplace, Shanghai) vs (X, birthplace, Beijing)
    \item Or cyclic/contradictory relational structure that cannot both be true under common-sense constraints. Example: (A, father, B) vs (B, father, A)
\end{itemize}

\textbf{2. Temporal conflict (type = ``temporal")}
A temporal conflict happens when:
\begin{itemize}
    \item The predicate describes a role/state that can change over time and is typically unique at a given moment
    (e.g., president/CEO/champion/current location).
    \item If both triples claim different objects for the same subject-predicate:
    \begin{itemize}
        \item If explicit time scopes exist and overlap $\rightarrow$ hard temporal conflict.
        \item If time scopes exist and do NOT overlap $\rightarrow$ not a conflict.
        \item If time scopes are missing but the predicate is time-variant and moment-unique $\rightarrow$ suspected temporal conflict
        (ask for time ranges; do NOT assert a hard conflict without time info).
    \end{itemize}
\end{itemize}

\textbf{3. Granularity conflict (type = ``granularity")} 
\begin{itemize}
    \item Triples differ due to specificity/abstraction level.\\
    \texttt{Example: (X, birthplace, Shanghai) vs (X, birthplace, China)}
    
    \item If one object is a parent/superset/contains the other (hypernym/meronym/administrative containment), then it is usually compatible $\rightarrow$ classify as \texttt{"granularity"}.
    
    \item If objects are incompatible (cannot contain each other and cannot both be true) $\rightarrow$ Logical conflict.
\end{itemize}

Output MUST be a valid JSON object following the required schema.
\end{mypromptbox}
\caption{The prompt used for Conflict Detection Agent.}
\label{appendix:prompt1}
\end{figure*}

\vspace{0.3cm}

\begin{figure*}[t!] 
\begin{mypromptbox}[Conflict Resolution]{promptGreen}
\textbf{Task Definitions:} You are an expert knowledge graph curator. Given a set of conflicting triples and their source passages, your task is to resolve the conflicts and produce corrected triples.

\textbf{Conflict Resolution Strategies:}

\textbf{1. Mutual Conflict (type = ``mutual"):} These are contradictory claims about the same entity (e.g., same subject-predicate but different objects)
\begin{itemize}
\item Resolution: Analyze the source passages to determine which triple is more accurate

\item Keep only the CORRECT triple, discard the incorrect one(s)

\item If both seem equally valid based on context, prefer the one with more specific/credible source

\end{itemize}

\textbf{2. Temporal Conflict (type = ``temporal"):} These are time-dependent facts where time scopes overlap or are missing
\begin{itemize}
\item Resolution: Add time information to the relation to distinguish the facts

\item Modify the predicate to include time context (e.g., "was president of [2000-2005]" vs "was president of [2005-2010]")

\item If time info is not in sources, note it as ``temporal\_conflict\_unresolved"

\end{itemize}

\textbf{3. Granularity Conflict (type = ``granularity"):} These are facts at different levels of specificity (e.g., "born in Shanghai" vs "born in China")
\begin{itemize}
\item Resolution: Add granularity description to the relation to clarify the scope

\item Modify the predicate to include granularity context (e.g., "was born in [city: Shanghai]" vs "was born in [country: China]")

\item Both can be kept if they are compatible (containment relationship)
\end{itemize}

Output MUST be a valid JSON object following the required schema.
\end{mypromptbox}
\caption{The prompt used for Conflict Resolution Agent.}
\label{appendix:prompt2}
\end{figure*}

\subsection{Memory-guided Online Retrieval}\label{appedix:retrieval_details}
Building upon the constructed Global Hierarchical Graph $\mathcal{G}$ and the Global Memory $\mathcal{M}$, this section details our memory-guided retrieval and reasoning mechanism. To bridge the gap between the user query and the complex graph topology, the inference workflow unfolds through three logically progressive stages: The workflow consists of three key steps:  i) \textbf{Multi-Layer Memory Retrieval}, which retrieves initial initial candidate evidence, including schemas $s$, facts $f$, and passages $p$ from $M_{ont}$, $M_{fac}$, and $M_{pas}$, respectively. It then applies a preliminary noise filtering process to ensure relevance. ii) \textbf{Structure-Aware Node Initialization}, which projects the retrieved evidence onto the graph structure by mapping them to initial node weights. We apply distinct scoring strategies for Entity nodes $e$, Type nodes $t$, and Passage nodes $p$, integrating semantic relevance, topological constraints, and information density. iii) \textbf{Graph Propagation}, which executes the Personalized PageRank (PPR) algorithm on the heterogeneous graph, initiating from the weighted nodes. This propagation diffuses importance across the graph to identify the most globally significant passages and nodes, which are then selected for downstream LLM generation.

\subsubsection{Multi-Layer Memory Filtering}
The retrieval phase initiates by querying the three distinct layers of the Global Memory$\mathcal{M}$ in parallel. Given a user query $\mathbf{q}$, we parallelly retrieve top-$K$ candidates from $M_{ont}$, $M_{fac}$, and $M_{pas}$ respectively. To prevent low-relevance noise from propagating into the graph reasoning stage, we apply a strict relevance filter. For the retrieved schemas $\mathcal{S}_{ret}$ and facts $\mathcal{F}_{ret}$, only candidates satisfying a semantic similarity threshold $\text{Sim}(\mathbf{q}, \mathbf{x}) > \tau$ are retained. This filtering ensures that the subsequent node initialization is seeded exclusively with high-confidence structural evidence. Crucially, to guarantee system robustness, if the filtering process yields no valid structural evidence (i.e., $S_{\mathrm{ret}} \cup F_{\mathrm{ret}} = \emptyset$), the framework adaptively falls back to a standard RAG mode, relying solely on the direct similarity between the query and the content in $M_{pas}$ for answer generation.

\subsubsection{Structure-Aware Node Initialization} 
To seed the subsequent graph propagation process with specific semantic context, we must project the retrieved evidence onto the heterogeneous graph topology. Formally, we define an initial reset probability distribution $P_{init}(v)$ for any node $v\in\mathcal{G}$. This distribution provides an initial importance score for the inference algorithm, quantifying the intrinsic significance of each node prior to information diffusion.

\textbf{1. Entity Node Initialization via Facts:} To ensure that graph propagation originates from grounded evidence, we first initialize entity nodes based on the relevance of their associated facts retrieved from $\mathcal{M}_{fac}$. Formally, we quantify the initial importance of an entity e as the mean semantic similarity of all filtered facts containing it:

\begin{equation}
    P_{init}(e) = \frac{1}{|\mathcal{F}_e|} \sum_{f \in \mathcal{F}_e} \text{Sim}(\mathbf{q}, \mathbf{f})
\end{equation}

where $\mathcal{F}_e \subseteq \mathcal{F}_{ret}$ denotes the subset of query-relevant facts contain entity $e$. If $\mathcal{F}_e=\emptyset$, the weight defaults to 0. This aggregation strategy ensures that entities are activated strictly by explicit, query-relevant factual support.

\textbf{2. Type Node Initialization via Schemas:} To incorporate macro-level domain knowledge and avoid introducing irrelevant semantics, we further initialize type nodes $t \in \mathcal{G}_{\text{schema}}$ based on the retrieved schemas from $\mathcal{M}_{\text{ont}}$. A critical challenge is that type nodes often exhibit disproportionately large degrees (e.g., a generic “Person” node connected to thousands of entities). Activating such high-degree nodes directly would cause importance to diffuse too broadly across the graph, thereby introducing substantial noise. To address this issue, we introduce a structural regularization term that combines semantic relevance with a log-degree penalty:

\begin{equation}
    P_{init}(t) = \underbrace{\left( \frac{1}{|\mathcal{S}_t|} \sum_{s \in \mathcal{S}_t} \text{Sim}(\mathbf{q}, \mathbf{s}) \right)}_{\text{Schema Relevance}} \times \underbrace{\frac{1}{\log(\text{deg}(t) + 1)}}_{\text{Hub Suppression}}
\end{equation}

where $\mathcal{S}_t$ denotes the subset of retrieved schemas corresponding to type $t$, $\text{deg}(t)$ is the node degree. This formulation effectively leverages ontology as a weak supervision signal while strictly constraining the diffusion radius of overly generic concepts.

\textbf{3. Passage Initialization with Information Density:} Finally, we need to initialize the Passage Nodes ($p \in G_{pas}$). We formulate the comprehensive scoring function to prioritize semantically relevant sources with high-value information, while avoiding dominance over finer-grained entity nodes, as follows:
\begin{equation}
    P_{init}(p) = \text{Sim}(\mathbf{q}, \mathbf{d}_p) \times \alpha \times \underbrace{\sigma\left( \frac{\sum_{e \in \mathcal{E}_p} \text{IDF}(e)}{\log(|\mathcal{E}_p| + 1)} \right)}_{\text{Information Density Term}}
\end{equation}
This formula integrates three critical dimensions: (i) \textit{Semantic Alignment} ($\text{Sim}$), which measures the vector similarity between the query $q$ and the passage embedding $d_p$; (ii) \textit{Structural Balance} ($\alpha$), a dampening coefficient empirically set to 0.05, which prevents dense passage nodes from overwhelming sparse entity nodes during the initial propagation phase and ensures a balanced importance distribution; and (iii) \textit{Information Density Term}, which quantifies content quality by summing the Inverse Document Frequency (IDF) of entities $E_p$ within the passage and applying log-normalization, thereby rewarding passages that contain rare and discriminative facts rather than generic, verbose content.

\subsubsection{Personalized PageRank}
Following the initialization phase, We execute the Personalized PageRank (PPR) algorithm on the heterogeneous graph to diffuse the initial semantic energy. The propagation uses the normalized vector $\mathbf{p}^{(0)}$ as the starting distribution and follows the iteration:
\begin{equation}
    \mathbf{v}^{(k+1)} = (1 - \lambda) \mathbf{W} \mathbf{v}^{(k)} + \lambda \mathbf{v}^{(0)}
\end{equation}
where $\mathbf{W}$ is the transition matrix of the graph. We specifically set the damping factor $\lambda = 0.5$ to restrict the random walk to a local neighborhood, thereby preventing semantic drift into irrelevant multi-hop connections. Upon convergence to $\mathbf{v}^{(\infty)}$ the top-K passages and top-M entities with the highest scores are selected as the context window for LLM inference.

\section{Benchmark Dataset}\label{app:dataset}
We first evaluate the effectiveness of MemGraphRAG on three widely-used multi-hop QA datasets, including HotpotQA \citep{yang2018hotpotqa}, 2WikiMultiHopQA (2Wiki) \citep{2wikimqa} and MuSiQue \citep{trivedi2022MuSiQue} and two GraphRAG benchmarks: G-bench (Novel) and G-bench (Medical) \citep{xiang2025use}. We provide a concise overview of each dataset's key characteristics below.

\textbf{(i) HotpotQA~\cite{yang2018hotpotqa}:} A widely adopted dataset for evaluating multi-hop reasoning across disparate texts. It requires models to filter through distractor paragraphs and synthesize information from multiple supporting documents to answer complex queries, thereby testing the system's ability to perform effective cross-document evidence retrieval.

\textbf{(ii) 2WikiMultiHopQA (2Wiki)~\cite{2wikimqa}:} A benchmark derived from Wikipedia knowledge graphs, specifically constructed to test structured reasoning. It consists of queries that necessitate aggregating evidence chains from two to four specific articles, focusing on the model's capacity to handle complex entity relationships and maintain logical consistency.

\textbf{(iii) MuSiQue~\cite{trivedi2022MuSiQue}:} A challenging dataset designed to minimize reasoning shortcuts often found in earlier benchmarks. It features connected reasoning chains of 2-4 hops, requiring systems to perform strictly sequential logical inference across multiple documents to derive the correct answer.

\textbf{(iv) G-bench (Novel) \& G-bench (Medical)}~\citep{xiang2025use}: Two domain-specific benchmarks tailored to evaluate GraphRAG performance on hierarchical retrieval and deep contextual understanding. The \textit{Medical} subset utilizes NCCN guidelines to test the handling of dense, rule-based clinical protocols, while the \textit{Novel} subset employs literary texts from Gutenberg to assess the comprehension of implicit, non-linear narrative structures.

\section{Implementation Details of Baselines}\label{app:baseline}
In our experiments, we compare our method against several widely used GraphRAG models.

\textbf{KGP}~\cite{wang2024knowledge} facilitates multi-document question answering by constructing a graph where nodes represent passages or document structures. It employs an LLM-driven traversal agent to navigate semantic and structural connections, progressively aggregating supporting context for the final response.

\textbf{G-Retriever}~\cite{he2024g} targets real-world textual graphs by formulating the subgraph retrieval task as a Prize-Collecting Steiner Tree (PCST) optimization problem. This approach extracts the most relevant subgraph to fit within the LLM context window, enabling effective conversational QA while mitigating hallucination and ensuring scalability.

\textbf{RAPTOR}~\cite{sarthi2024raptor} employs a recursive abstraction approach to construct a hierarchical tree structure. By clustering and summarizing text chunks from the bottom up, it enables the retrieval of information at varying levels of granularity, capturing both high-level context and fine-grained details for holistic understanding.

\textbf{MS-GraphRAG}~\cite{edge2024local} enhances global corpus understanding by building an entity-relation graph and pre-computing community-level summaries. It answers queries by synthesizing insights from these communities, offering improved comprehensiveness for questions that span the entire document collection.

\textbf{LazyGraphRAG}~\cite{Lazygraphrag} introduces a cost-effective paradigm that eliminates the need for expensive up-front summarization of source data. By avoiding the pre-computation of community hierarchies, it reduces indexing costs to the level of standard vector RAG while maintaining superior performance on local queries and competitive quality on global queries compared to full-graph approaches.

\textbf{LightRAG}~\cite{guo2024lightrag} introduces a two-tier retrieval strategy designed to capture both detailed entity relationships and broader thematic contexts. It utilizes graph-enhanced indexing to facilitate rapid access to relevant information and allows for seamless integration of new data via an incremental update algorithm.

\textbf{HippoRAG}~\cite{hipporag} proposes a neurobiologically inspired framework that orchestrates LLMs, knowledge graphs, and Personalized PageRank. It acts as a dual-system memory model to enable deep knowledge integration, facilitating robust retrieval for scenarios requiring the synthesis of information from multiple sources.

\textbf{HippoRAG2}~\cite{gutiérrez2025hipporag2} extends the Personalized PageRank-based framework of its predecessor by optimizing passage contextualization and the online interaction with LLMs. These enhancements enable the model to mimic human long-term memory more effectively, balancing robust factual recall with complex associative reasoning.

\textbf{E$^2$GraphRAG}~\cite{zhao20252graphrag} optimizes the GraphRAG paradigm by establishing bidirectional indexes between document chunks and entities. It combines a summary tree with a lightweight entity graph to facilitate fast lookup, enabling an adaptive retrieval process that seamlessly integrates local context and global understanding without manual query mode selection.

\textbf{GFM-RAG}~\cite{luo2025gfm} introduces a Graph Foundation Model (GFM) designed for zero-shot application on unseen datasets. It employs a pre-trained Graph Neural Network to reason over graph structures, effectively capturing complex query-knowledge relationships while mitigating the impact of noise and incompleteness in the constructed graphs.

\textbf{LogicRAG}~\cite{chen2025logicrag} introduces a dynamic retrieval paradigm where query-specific logic is modeled as a directed acyclic graph at inference time. By linearizing this graph via topological sort, it guides the retrieval process through a logically consistent sequence of sub-problems, significantly reducing token usage compared to static graph approaches.

\textbf{LinearRAG}~\cite{zhuang2025linearrag} challenges the reliance on costly and unstable relation extraction in existing methods. It constructs a relation-free hierarchical structure termed ``Tri-Graph'' using lightweight entity extraction and semantic linking. This approach scales linearly with corpus size and employs a two-stage retrieval strategy involving local entity activation and global importance aggregation.

\end{document}